\documentclass[aps, prd, 10pt, twocolumn, nofootinbib, superscriptaddress,noshowpacs, preprintnumbers, longbibliography, hyperref,floatfix]{revtex4-1}

\pdfoutput=1
\usepackage[title]{appendix}
\usepackage[colorlinks=true,breaklinks=true]{hyperref}
\usepackage[normalem]{ulem}
\usepackage[utf8]{inputenc}
\hypersetup{allcolors=[rgb]{0.0 0.0 0.6},linkcolor=[rgb]{0.75 0.05 0.05}}
\usepackage{amsmath,amssymb}
\usepackage{epsfig}  
\usepackage{graphicx}   
\usepackage{url}
\usepackage[dvipsnames]{xcolor}
\usepackage{color}
\usepackage{multirow}
\usepackage{comment}
\usepackage{amssymb}
\usepackage[version=3]{mhchem}
\usepackage{orcidlink}
\usepackage{longtable}
\usepackage{printlen}

\DeclareMathOperator{\GeV}{GeV}

\DeclareMathOperator{\kpc}{kpc}

\DeclareUnicodeCharacter{2212}{-}

\allowdisplaybreaks

\bibpunct{[}{]}{,}{n}{}{,}
\definecolor{darkspringgreen}{rgb}{0.09, 0.45, 0.27}

\makeatletter
\renewcommand\section{\@startsection{section}{1}{\z@}%
  {-2.5ex \@plus -1ex \@minus -.2ex}
  {1.2ex \@plus .2ex}
  {\reset@font\normalsize\bfseries\centering}} 
\makeatother

\begin{document}

\title{Detecting light axions from supernovae in  nearby galaxies}

\author{Francesca Lecce~\orcidlink{0009-0009-6967-5042}}
\email{francesca.lecce@ba.infn.it}
\affiliation{Dipartimento Interuniversitario di Fisica  ``Michelangelo Merlin'', Via Amendola 173, 70126 Bari, Italy}
\affiliation{Istituto Nazionale di Fisica Nucleare - Sezione di Bari, Via Orabona 4, 70126 Bari, Italy}%

\author{Alessandro Lella~\orcidlink{0000-0002-3266-3154}}
\email{alessandro.lella@ba.infn.it}
\affiliation{Dipartimento Interuniversitario di Fisica  ``Michelangelo Merlin'', Via Amendola 173, 70126 Bari, Italy}
\affiliation{Istituto Nazionale di Fisica Nucleare - Sezione di Bari, Via Orabona 4, 70126 Bari, Italy}%

\author{Giuseppe Lucente~\orcidlink{0000-0002-3266-3154}}
\email{lucenteg@slac.stanford.edu}
\affiliation{SLAC National Accelerator Laboratory, 2575 Sand Hill Rd, Menlo Park, CA 94025}

\author{Maurizio Giannotti~\orcidlink{0000-0001-9823-6262}}
\email{mgiannotti@unizar.es}
\affiliation{Centro de Astropartículas y Física de Altas Energías, University of Zaragoza, Zaragoza, 50009, Aragón, Spain}
\affiliation{Physical Sciences, Barry University, 11300 NE 2nd Ave., Miami Shores, FL 33161, USA}

\author{Alessandro Mirizzi~\orcidlink{0000-0002-5382-3786}}
\email{alessandro.mirizzi@ba.infn.it}
\affiliation{Dipartimento Interuniversitario di Fisica  ``Michelangelo Merlin'', Via Amendola 173, 70126 Bari, Italy}
\affiliation{Istituto Nazionale di Fisica Nucleare - Sezione di Bari, Via Orabona 4, 70126 Bari, Italy}%

\begin{abstract}
Axion-like particles (ALPs) coupled to nucleons can be efficiently  produced in core-collapse supernovae (SNe) and then,  if they couple to photons,  convert into gamma rays in cosmic magnetic fields, generating short gamma-ray bursts. Though ALPs from a Galactic SN would induce an intense and easily detectable gamma-ray signal, such events are exceedingly rare. In contrast, a few SNe per year are expected {in nearby galaxies} within $\sim \mathcal{O}(10)$ Mpc, where strong magnetic fields can enable more efficient ALP–photon conversions than in the Milky Way, offering a promising extragalactic target. 
This circumstance motivates full-sky gamma-ray monitoring, ideally combined with deci-hertz gravitational-wave detectors to enable time-triggered searches from nearby galaxies. We show that, under realistic conditions, a decade of coverage could reach sensitivities to the product of the
ALP-proton and ALP-photon couplings $g_{ap}\times g_{a\gamma} \gtrsim 10^{-24}~{\rm GeV}^{-1}$ for ALP masses $m_a \lesssim 10^{-9}$~eV. This sensitivity would allow one to
probe a large, currently-unexplored region of the parameter space below the longstanding SN 1987A bound.

\end{abstract}

\date{\today}
\smallskip

\maketitle

\section{Introduction}
\begin{figure} [t!]
\centering
\includegraphics[width=1\columnwidth]{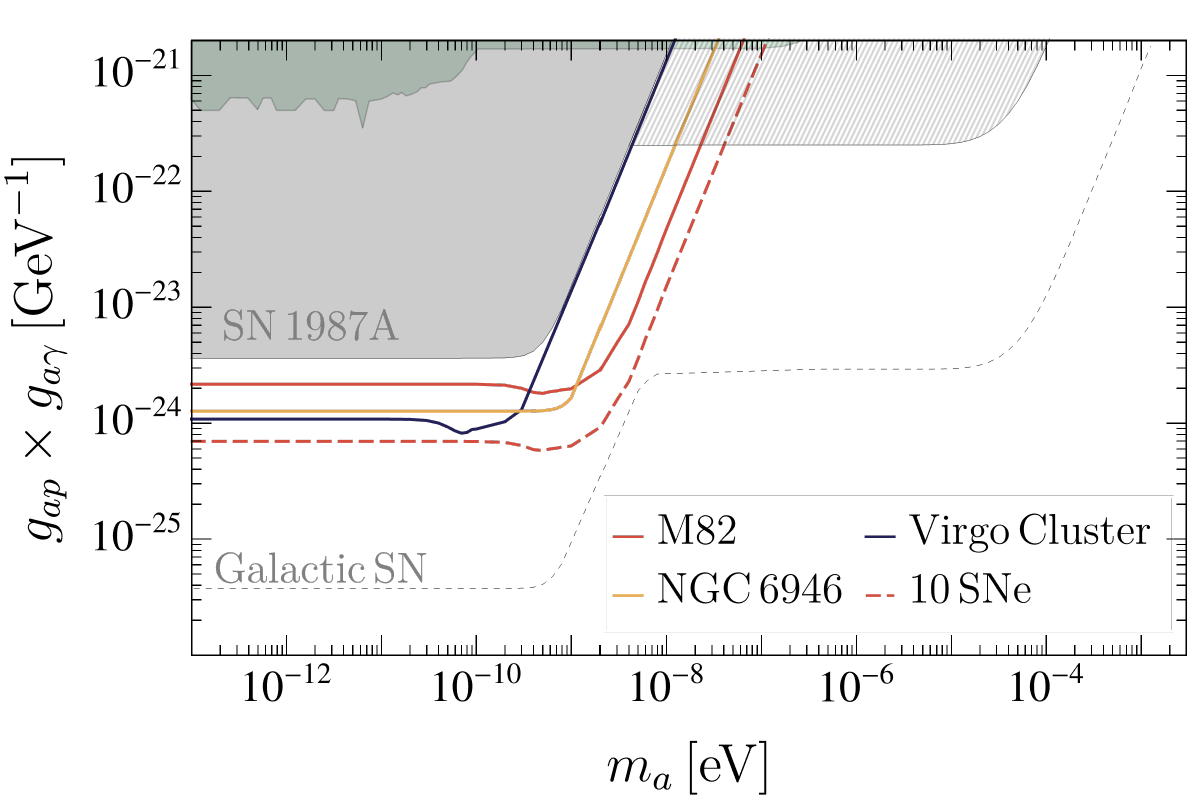}
\caption{Sensitivities {at $95\%$ CL} in the $g_{ap}\times g_{a\gamma}$ \emph{vs} $m_a$ plane for future extragalactic SN occurring in M82 (red line), NGC 6946 (yellow line) or Virgo Cluster (black line) for a {\emph{Fermi}-LAT-like} gamma-ray detector, assuming an observation time window of $\Delta t=10$~s enabled by the core-collapse time trigger provided by deci-hertz GW interferometers. 
    The dashed red curve illustrates the stacked sensitivity from 10 SNe in an environment similar to M82. The gray band displays the constraint from the non-observation of gamma rays from SN 1987A for ALPs converting in the Milky-Way, while the hatched region would be probed by conversions in the SN 1987A progenitor magnetosphere assuming a $100\,\rm G$ surface field~\cite{Manzari:2024jns}, but no measurements of such a field are available~\cite{Fiorillo:2025gnd}. For comparison, we also show the sensitivity to a future Galactic SN at a distance of $d=10$~kpc (dashed black curve) by assuming the same magnetic field models used for the SN 1987A limit. Finally, we show in green the dominant astrophysical bounds in this mass range~\cite{AxionLimits,Reynes:2021bpe,Ning:2024eky,Benabou:2025jcv}.}
    \label{fig:fig1}
\end{figure}

Axion-like particles (ALPs) 
emerge in several extensions of the Standard Model of particle physics~\cite{Jaeckel:2010ni,Ringwald:2014vqa,DiLuzio:2020wdo,Agrawal:2021dbo,Giannotti:2022euq,Antel:2023hkf,Arza:2026rsl}. 
From a top-down perspective, string theory predicts the presence of an ``axiverse'' with the QCD axion~\cite{Peccei:1977hh,Peccei:1977ur,Weinberg:1977ma,Wilczek:1977pj} and several ultralight ALPs~\cite{Arvanitaki:2009fg,Cicoli:2012sz,Cicoli:2023opf}.
From a bottom-up perspective, ALPs
offer an interesting physics case in relation to dark matter~\cite{Abbott:1982af,Dine:1982ah,Preskill:1982cy,Arias:2012az,Adams:2022pbo} and 
several astrophysical puzzles~\cite{Giannotti:2015kwo,Giannotti:2017hny,DiLuzio:2021ysg,Galanti:2022chk}.
In this context, stars provide ideal environments for ALP production due to their hot, dense plasmas~\cite{Raffelt:1996wa,Raffelt:1999tx} (see Refs.~\cite{Caputo:2024oqc,Carenza:2024ehj,Arza:2026rsl} for recent reviews). The Sun, being the closest star to Earth, has been extensively studied using ``helioscope'' experiments~\cite{vanBibber:1988ge,IAXO:2019mpb,IAXO:2020wwp}  and X-ray detectors~\cite{Ruz:2024gkl}. 
Nevertheless, other stellar systems offer competitive discovery potential~\cite{Carenza:2024ehj}.
Remarkably, following the detection of neutrinos from Supernova SN 1987A~\cite{Kamiokande-II:1987idp,Hirata:1988ad,Alekseev:1988gp,Bionta:1987qt,IMB:1988suc}, it was proposed that ALPs could be copiously produced in the supernova (SN) core alongside neutrinos ~\cite{Brinkmann:1988vi,Burrows:1988ah}. In particular, a substantial ALP flux can be generated via interactions with nucleons in the hot and dense SN core~\cite{Brinkmann:1988vi,Keil:1996ju,Carenza:2019pxu,Carenza:2020cis}. If also coupled with photons, a fraction of these ALPs could convert into gamma rays in the Milky Way magnetic field, yielding a short gamma-ray burst coincident with the neutrino signal. The absence of such a signal in the Gamma-Ray Spectrometer aboard the Solar Maximum Mission led to stringent constraints on the product of the ALP-proton and ALP-photon couplings~\cite{Grifols:1996id,Brockway:1996yr,Payez:2014xsa,Hoof:2022xbe,Manzari:2024jns,Fiorillo:2025gnd}, excluding ${g_{ap} \times g_{a\gamma} \gtrsim {3.6} \times 10^{-24}}$~GeV$^{-1}$ for $m_a \lesssim 10^{-10}$~eV (see the gray band in Fig.~\ref{fig:fig1}).

The observation of a future Galactic core-collapse supernova (CCSN) would enable tests of substantially smaller ALP couplings~\cite{Meyer:2016wrm,Calore:2023srn,Lella:2024hfk}.
For example, {if the \emph{Fermi} Large Area Telescope (LAT) were to observe such a SN event, it}
could probe down to ${g_{ap} \times g_{a\gamma} \gtrsim {3.7} \times 10^{-26}}$~GeV$^{-1}$ for ultra-light ALPs with $m_a \ll 10^{-9}$~eV (see the dashed black curve in Fig.~\ref{fig:fig1}).
In addition, including ALP–photon conversions in the strong magnetic field of the progenitor star may even grant access to the canonical QCD axion parameter space~\cite{Manzari:2024jns,Fiorillo:2025gnd,Candon:2025sdm}. 
Though all this sounds extremely promising, Galactic SNe are rare events, occurring at a rate of only about two per century~\cite{Rozwadowska:2020nab}.
A different approach consists in the observation of individual extragalactic SNe, 
producing ALPs that convert in gamma-rays within the Milky Way magnetic field.
This idea  was first explored in Ref.~\cite{Meyer:2020vzy}, which investigated the sensitivity of {\it Fermi}-LAT to  extragalactic events, at a distance $d \lesssim270$~Mpc. 
However, unlike Galactic SNe, the core-collapse time cannot be precisely determined because neutrino bursts from such distances are undetectable with current and near-future underground experiments~\cite{Mukhopadhyay:2021gox}. In the absence of a neutrino trigger, the explosion time must  be inferred from the optical light curves~\cite{Heston:2023gbx}. This substantially enlarges the observational time window during which ALP-induced gamma rays may appear, transforming an otherwise background-free search into a background-dominated one. Additionally, the limited field of view of {\it Fermi}-LAT may prevent coverage at the relevant core-collapse time~(see Ref.~\cite{Candon:2025ypl} for a discussion of how the probability that Fermi-LAT observes at least one SN at the time of collapse scales with the sample size).
For this reason, extragalactic SNe have remained largely off the radar.
The aim of this article is to show that this picture can change dramatically if forthcoming experimental proposals are realized. Furthermore, nearby galaxies may host magnetic environments in which the probabilities of ALP-photon conversion are enhanced by 2–3 orders of magnitude relative to the Milky Way~\cite{Ning:2024eky,Ning:2025tit,Ning:2025kyu}.  With the proper experimental setup, extragalactic SNe could become a genuinely transformative tool, allowing us to probe ALP parameter space far beyond current limits (see Fig.~\ref{fig:fig1}).
\begin{figure} [t!]
\centering
 \includegraphics[width=1\columnwidth]{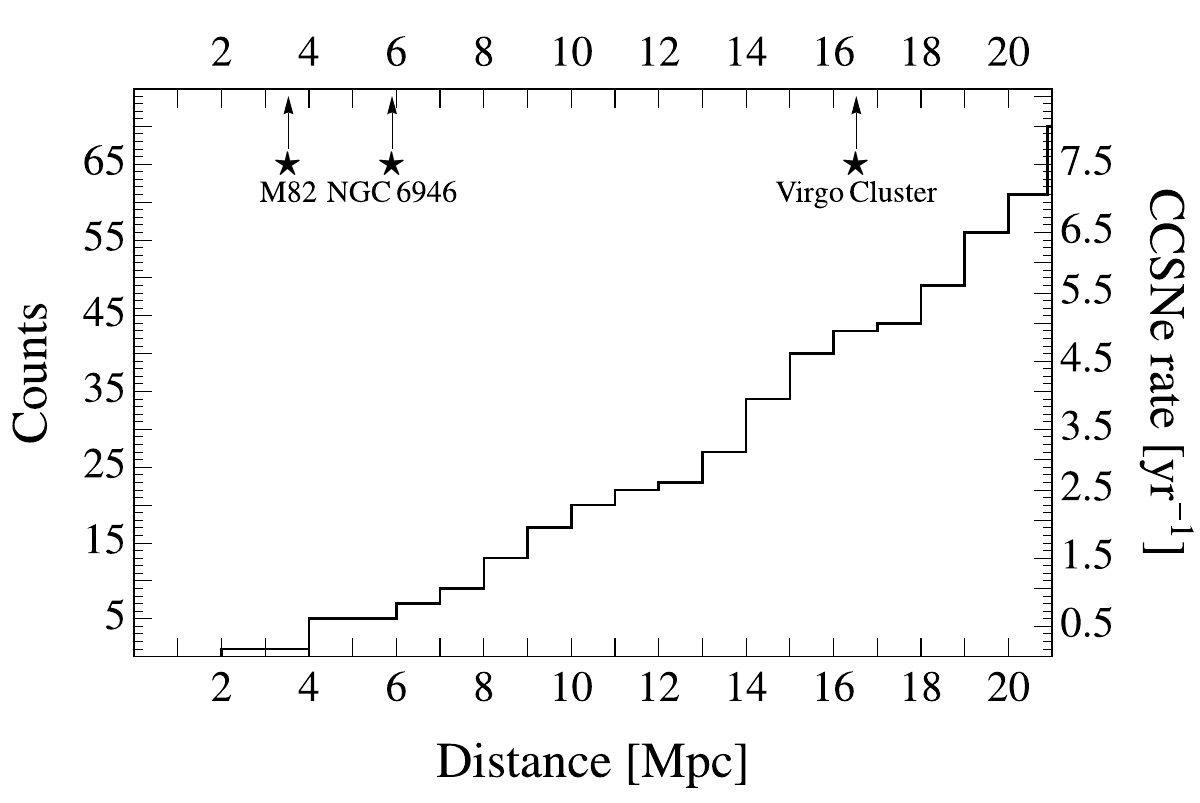}
    \caption{Cumulative number of observed core-collapse supernovae in the period 2008-2025 (left vertical axis) from nearby galaxies in function of the distance. On the right vertical axis is also indicated the inferred SN rate per year. }
	\label{fig:fig2}
\end{figure}
The paper is structured as follows. In Sec.~\ref{sec:Whatisneeded?} we evaluate the rate of extragalactic SNe and discuss the prerequisites to exploit these sources. In Sec.~\ref{sec:ALP production in SN core.} we discuss ALP production from $NN$ bremsstrahlung. Sec.~\ref{sec:ALP-photon conversions in host galaxies.} presents the typical conversion probabilities in host galaxies. In Sec.~\ref{sec:Sensitivities}, we show the obtained sensitivities. Finally, in Sec.~\ref{sec:Conclusions}, we summarize our results and draw our conclusions.

\section{ALPs from extragalactic SNe: What is needed?}

\label{sec:Whatisneeded?}
To  {clarify} the physics case for ALP searches from extragalactic SNe, in Fig.~\ref{fig:fig2} we plot the number of observed  CCSN rate up to a distance $d = 20$~Mpc reported in 
the Transient Name Server~\cite{2021AAS...23742305G} 
for the period 2008-2025, corresponding to the lifetime of {\it Fermi}-LAT. The detailed list of core-collapse SNe with their distances is reported in Sec.~A in the Supplemental Material(SM)~\cite{suppmat}. {From such a plot, one infers that roughly seven core-collapse SNe per year occurred within this volume.} 
{Therefore, extending the SN horizon to nearby galaxies dramatically increases the number of potential events, removing the dependence on the rare occurrence of a Galactic SN.}
Based on these results, in this article, we show that, to turn extragalactic CCSNe into powerful ALP laboratories, two key experimental capabilities are required:
\emph{(a)} a full-sky, continuously operating constellation of gamma-ray detectors and \emph{(b)} an external time trigger that determines the SN core-collapse time with sufficient accuracy.
These are not speculative or serendipitous conditions—they represent clear, achievable technological objectives whose realization would decisively enable extragalactic ALP astrophysics.
The first objective, condition \emph{(a)}, is well within reach. Concepts such as the GALAXIS mission~\cite{Manzari:2024jns} already outline a satellite constellation with Fermi-LAT–like sensitivity and genuine full-sky coverage. 
Although primarily designed for Galactic SNe, we show below that its physics reach extends naturally to extragalactic distances when combined with requirement \emph{(b)}.

This latter prerequisite is equally realistic. 
While large underground neutrino detectors are intrinsically limited to probing SNe within a few Mpc~\cite{Mukhopadhyay:2021gox},
current optical surveys, {which aim to discover SNe up to tens of Mpc may determine the core-collapse time with an accuracy of $\Delta t=1$~d~\cite{Ando:2005ka,Tartaglia:2017yym}.}  
Furthermore, future optical surveys, 
may reach
$\Delta t=1$~h~\cite{Heston:2023gbx}. Indeed, by modeling the early light curve of the SN,  Ref.~\cite{Cowen_2010} determined the core-collapse time of real SNe within an accuracy of better than a few hours.
Finally, next-generation deci-hertz GW observatories (DECIGO, BBO)~\cite{Seto:2001qf,Yagi:2011wg,Kawamura:2020pcg} are expected to detect the supernova gravitational \emph{neutrino-memory} signal out to tens of Mpc~\cite{Mukhopadhyay:2021zbt,Mukhopadhyay:2021gox}, providing the optimal explosion-time trigger needed for ALP searches. The memory develops $0.1\,$s from the onset of neutrino emission, making it an ideal timing marker. Building on these possibilities, we show that the presence of an accurate external trigger with low-latency alert dissemination may open new regions of ALP parameter space beyond those excluded by SN 1987A observations for nearby ($d\lesssim 20\,\rm{Mpc}$) extragalactic SNe.

\section{ALP production in SN core}

\label{sec:ALP production in SN core.}
The production of ALPs in the SN core has been extensively studied over the past decades~\cite{Carena:1988kr,Brinkmann:1988vi,Raffelt:1993ix,Raffelt:1996wa,Carenza:2019pxu,Lella:2022uwi,Lella:2023bfb,Fiorillo:2025gnd}. 
A particularly efficient production channel is nucleon–nucleon (NN) bremsstrahlung~\cite{Carenza:2019pxu}, mediated by the ALP–nucleon interaction described by the Lagrangian {${\mathcal L}_{aN}=
{(2m_N)^{-1}}{\partial_\mu a}\sum_{N=p,n}
g_{aN}{\bar N}\gamma^\mu\gamma_5 N$}, where $a$ is the ALP field, $m_N$ is the nucleon mass, and $g_{aN}$ is the ALP coupling to nucleon species, with $N=p,n$ for protons and neutrons, respectively. 
In the following, for simplicity, we assume ALPs coupling only to protons. In this work we calculate the SN ALP fluxes using as benchmark the 1D spherical symmetric {\tt GARCHING} group's SN model SFHo-s18.8 provided in Ref.~\cite{SNarchive} and based on the neutrino-hydrodynamics code 
{\tt PROMETHEUS-VERTEX}~\cite{Rampp:2002bq}. 
Notably, the NN bremsstrahlung produces  an ALP thermal spectrum with average energy {$E_a \simeq  70$ MeV}. The emissivity can be further enhanced if a sizable population of thermal pions is present in the SN core, enabling additional ALP production through pionic Compton-like $(\pi N)$ processes~\cite{Carenza:2020cis}. Because of the uncertainties in this process, we neglect it in the following discussion. 
See Sec.~B of the SM~\cite{suppmat} for more details on the SN ALP emissivity.

\section{ALP-photon conversions in host galaxies}

\label{sec:ALP-photon conversions in host galaxies.}
{Generically, ALPs can also couple}
to photons through the  Lagrangian term~\cite{Raffelt:1987yu,Raffelt:1987im},
 $ \mathcal{L}_{a\gamma} 
 =g_{a\gamma} a \, \mathbf{E} \cdot \mathbf{B}$,
where $g_{a\gamma}$ is the ALP-photon coupling.  
In this scenario, 
ultralight ALPs would efficiently convert within the cosmic magnetic fields, giving rise to a gamma-ray signal observable by gamma-ray experiments operating in the MeV energy range.
In Ref.~\cite{Meyer:2020vzy} the conversions of ALPs from extragalactic SNe were characterized by assuming only the presence of the Milky Way magnetic field. Therefore, compared to the case of a Galactic SN at $d\sim10$~kpc, we expect a flux penalty factor for extragalactic SNe  {${\mathcal O} ([10 \,\ \textrm{kpc}/10 \,\ \textrm {Mpc}]^2) \sim 10^{-6}$}, leading to a reduction in the sensitivity to $g_{ap}\times g_{a\gamma}$ of three orders of magnitudes.
However, the penalty associated with the much larger distances of extragalactic SNe can be substantially mitigated by the presence of strong magnetic fields in and around their host galaxies. As recently shown in Ref.~\cite{Ning:2024eky}, ALPs produced in stellar populations of nearby galaxies can undergo highly efficient conversions in these environments.
A particularly striking example is the M82 starburst galaxy ($d = 3.5$~Mpc), where magnetic fields enhanced by starburst activity can reach $\sim$0.1–1 mG~\cite{Lopez_Rodriguez_2021}, leading to sizable ALP--photon conversion probabilities~\cite{Ning:2024eky,Ning:2025tit,Ning:2025kyu}.\,\footnote{{Observations of this galaxy also allow constraints on massive ALPs decaying into photons~\cite{Candon:2024eah}.}}\,
Ref.~\cite{Ning:2024eky} also examined the elliptical galaxy M87 ($d=16.9$~Mpc). 
The latter would benefit from its position inside the Virgo Cluster, where Mpc-scale intracluster magnetic fields provide long conversion baselines that can compensate the inherently weaker strengths.
Following Ref.~\cite{Ning:2024eky}, we model the $B$-field and the electron density in these environments using the cosmological magnetohydrodynamical simulations  from \textsc{IllustrisTNG}~\cite{Nelson:2019jkf,Marinacci:2017wew},  as documented in Sec.~C of the SM~\cite{suppmat}.
As an additional example, we consider a third system, the spiral galaxy NGC 6946 at $d = 5.9$~Mpc. 
Although relatively small ($L \sim 9$~kpc), it hosts magnetic fields of the order of ${\mathcal O}(10)$~$\mu$G~\cite{Khademi_2023}.
Across all three environments, we compared the TNG-based predictions with existing models in the literature. The resulting conversion probabilities are consistent with, or even exceed, those derived from \textsc{IllustrisTNG}, demonstrating the robustness of our approach (see Sec.~C of the SM~\cite{suppmat} for further details about the computation of conversion probabilities).

In order to calculate ALP-photon conversions, we closely follow the technique described in Ref.~\cite{Horns:2012kw}, to which we address the reader for more details. For each subhalo modeling the properties of the environment, we compute the conversion probabilities across $10^4$ realizations, sampled by varying $10^2$ CCSN positions and $10^2$ observation directions. We define our \emph{fiducial} configuration as the one yielding the median conversion probability.
This latter for $g_{a\gamma}=10^{-12}\,\GeV^{-1}$ assumes values 
$P_{a\gamma}=8.8 \times 10^{-4}$ for M82,
$P_{a\gamma}=6.8 \times 10^{-3}$ for NGC 6946, and $P_{a\gamma}=7.6 \times 10^{-2}$ for { Virgo Cluster},  in the limit of $m_a \lesssim 10^{-9}$~eV. The observed gamma-ray flux will scale as $\Phi_{\gamma} \propto P_{a\gamma}/d^2$, implying  $\Phi_{\gamma}^{\rm M82}:\Phi_{\gamma}^{\rm NGC \,\  6946}: \Phi_{\gamma}^{\rm M87}=1:2.7:4.0$.  Therefore, despite the differences in the input parameters, one expects a similar sensitivity to $g_{a p} \times g_{a\gamma}$ as shown in Fig.~\ref{fig:fig1}.
Furthermore, we highlight that typical values of the conversion probability in the Milky Way are $P_{a\gamma}\sim{\mathcal O}(10^{-5})$ for $g_{a\gamma}=10^{-12}\,\rm{GeV}^{-1}$ and so negligible with respect to the values reached in all the considered  environments.
{ Indeed, accounting for conversions only in the Milky Way, for a SN exploding in M82  we would find a sensitivity to $g_{ap} \times g_{a\gamma}\sim {1.3\times 10^{-23}\,\rm{GeV}^{-1}}$, which is above the existing bound from SN 1987A.}
We also observe that the effect of a potentially strong magnetic field in the progenitor of the SN, as considered in~\cite{Manzari:2024jns,Fiorillo:2025gnd,Candon:2025sdm}, would extend the sensitivity to larger masses but in a region of the parameter space already excluded by other astrophysical constraints. Therefore, for simplicity we neglect also this effect.

\section{Sensitivities in $\boldsymbol{{g_{ap} \times g_{a\gamma}}}$}
\label{sec:Sensitivities}

\begin{figure} [t!]
\centering
    \includegraphics[width=1\columnwidth]{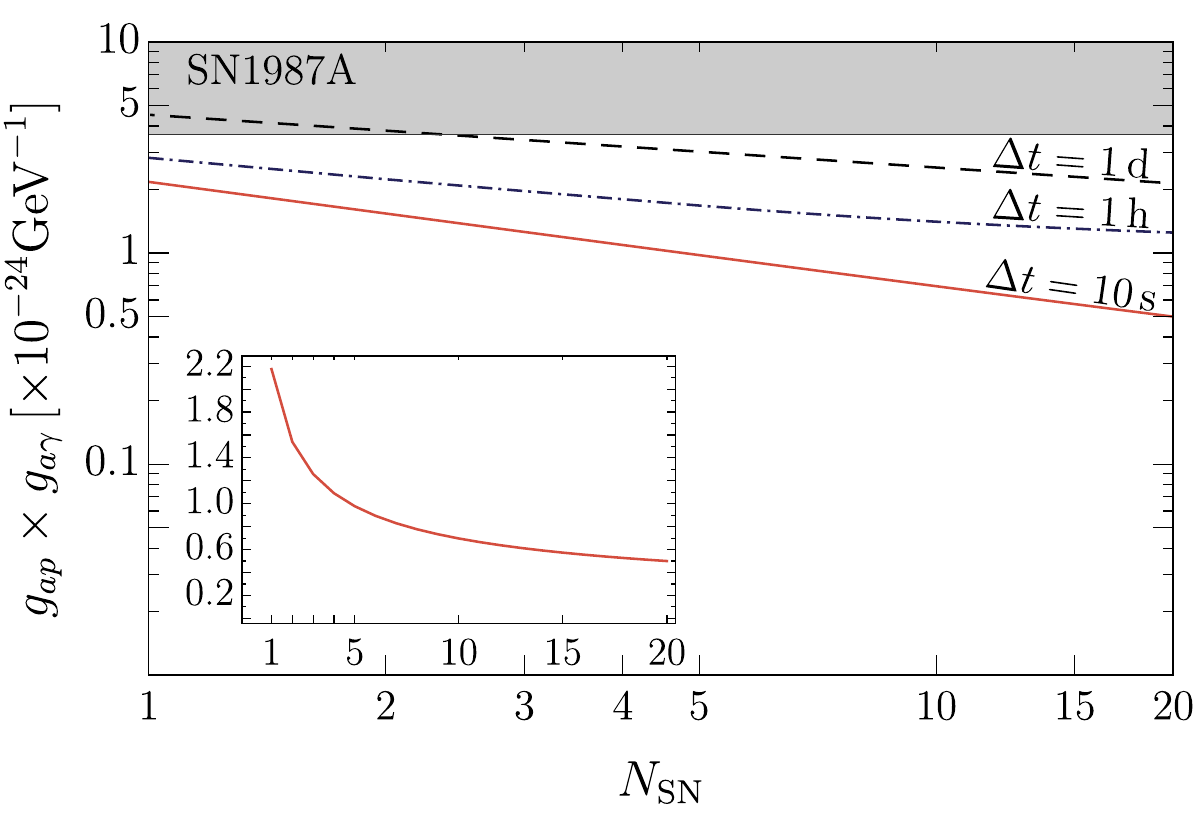}
    \caption{Sensitivity ($95\%$ CL) to  $g_{ap} \times g_{a\gamma}$ as a function of the number of observed extragalactic SNe, $N_{\rm SN}$, yielding a gamma-ray flux similar to the one estimated in the case of M82. We assume different time windows to search for the ALP signal, namely, $\Delta t=1$~d (dashed curve), $\Delta t=1$~h (dot-dashed curve), achievable with current and future optical surveys, and $\Delta t=10$~s (continuous curve), relying on an external trigger and operational conditions specified in the main text. In the inset, we show on a linear scale the sensitivity for $\Delta t = 10$~s scaling as $N_{\rm SN}^{-1/2}$. The $x$ and $y$ axes are the same as in the main panel.
    }
	\label{fig:fig4}
\end{figure}

With these results in hand, we are ready to derive the sensitivity to the product of ALP–nucleon and ALP–photon couplings $g_{ap} \times g_{a\gamma}$. 
Since the three nearby galaxies introduced before are expected to give a similar sensitivity in the low-mass ALP range ($m_a\lesssim 10^{-9}$~eV), for definiteness we focus on M82.
We assume to have a gamma-ray detector like the proposed project GALAXIS~\cite{Manzari:2024jns} with the same performances of {\it Fermi}-LAT, guaranteeing full sky coverage (see Sec.~D of the SM~\cite{suppmat}). 
In Fig.~\ref{fig:fig4}, we show 
the $95\%$ confidence level (CL) sensitivity~\cite{ParticleDataGroup:2024cfk} to $g_{ap} \times g_{a\gamma}$ in case of low-mass ALPs assuming different observation-time windows, determined by the accuracy in the determination of the SN {core-collapse} time. 
In low-background regime, the statistical analysis follows the Feldman–Cousins prescription~\cite{Feldman:1997qc}.

As a first case, we consider an accuracy 
of $\Delta t=1$~d (dashed curve) {determined by the extrapolation of the observed lightcurve} from   optical surveys ~\cite{Ando:2005ka,Tartaglia:2017yym}.  
In such a large observational time window, the gamma-ray detector is expected to accumulate a significant number of background events [$N_{\rm bkg} \sim{\mathcal O}(100)$]. Therefore, the statistical analysis points out that
if a SN explodes in M82, we will be sensitive to $g_{ap} \times g_{a\gamma} \gtrsim {{4.5}} \times 10^{-24}\,\rm{GeV}^{-1}$ at 95\% CL, which is a factor of {1.25} above the SN 1987A bound.
In this case, the significant improvement expected with respect to the results of Ref.~\cite{Meyer:2020vzy} is due to enhancement of the conversions in the magnetic environment of the host galaxy with respect to the Milky-Way, as shown in Sec.~F in the SM~\cite{suppmat}.

In order to further improve the sensitivities, it is crucial to reduce the background shortening the uncertainty on the determination of the SN core-collapse time. 
Let us first assume that future optical surveys might be able to reach 
$\Delta t=1$~h~(dot-dashed curve), as considered in Ref.~\cite{Heston:2023gbx}. 
With this accuracy, we get a sensitivity of {$g_{ap} \times g_{a\gamma} \gtrsim  {2.8} \times 10^{-24}\,\rm{GeV}^{-1}$, improving by a factor of $\sim {1.6}$ the case with $\Delta t=1$~d}. 
Finally, we consider the optimistic scenario of $\Delta t=10$~s (continuous curve) corresponding to the duration of the ALP burst. Achieving such a narrow observational window requires a precise determination of the burst onset, expected to be possible in the presence of a deci-Hertz GW interferometer observing the SN neutrino memory signal. This short-window scenario relies on specific operational conditions: \emph{(a)} an external trigger with $\mathcal{O}(10)\,s$ timing uncertainty, \emph{(b)} low latency in alert dissemination, and \emph{(c)} a nearby SN ($d\lesssim 20 \,\rm{Mpc}$) to ensure a high signal-to-noise ratio. At these distances, the trigger is robust as the stochastic GW background from cosmological SNe is negligible~\cite{Mukhopadhyay:2021gox}. 
In this case, since we are in a {background-free} regime, we compute the $95\%$~CL sensitivity by requiring that $\sim 3$ ALP-induced events are observed~\cite{Feldman:1997qc}, leading to $g_{ap} \times g_{a\gamma} \gtrsim {{2.3}\times 10^{-24}}\,\rm{GeV}^{-1}$. 
This corresponds to an improvement over the SN 1987A bound by a factor of $\sim {1.6}$.

We also estimate how the sensitivity would improve if multiple extragalactic SN explosions are observed over the detector’s lifetime. From the SN catalog covering the period 2008–2025 (Table~$\rm{A}1$ in Sec.~A in the SM~\cite{suppmat}), we find that seven SNe occurred during this time window in the three {environments} considered in this work (see Table~$\rm{A}3$). Since the expected ALP-induced signal from each of these systems is comparable to that of M82, observing seven such events is effectively equivalent to detecting seven explosions from M82 itself. Therefore, we show in Fig.~\ref{fig:fig4} the resulting sensitivity enhancement as a function of the number of observed SNe $N_{\rm SN}$, adopting M82 as a benchmark case. 
In a background-dominated situation, corresponding to SN core-collapse time known with an accuracy of $\Delta t=1$~d, the sensitivity to $g_{ap}\times g_{a\gamma}$ scales as $N_{\rm SN}^{-1/4}$.  
For $\Delta t = 1$ h, one finds
an intermediate regime between background-free and high-background scenarios.

Short coincidence windows, $\Delta t\approx 10$~s, enabled by a GW trigger, suppress the expected background and can place the search in the background-free regime.
Besides the obvious advantages of a reduced background, in this limit stacking becomes maximally efficient, with the coupling sensitivity scaling as $g_{ap}\, g_{a\gamma}\propto N_{\rm SN}^{-1/2}$.
Thus, reducing the time window directly accelerates the gain from multiple events. This provides strong motivation for synergies with next-generation GW detectors, which can sharply localize the burst time.
As an illustration, stacking $N_{\rm SN}=5$ CCSNe within $d\simeq 20$ Mpc, each with an M82-like expected flux, yields a projected sensitivity of
$g_{ap}\times g_{a\gamma}\simeq 9.8\times 10^{-25}\,\mathrm{GeV}^{-1}$,
an improvement of a factor of $3.7$ over the SN 1987A bound.
Remarkably, this corresponds to $g_{a\gamma} \gtrsim 9.8\,\times 10^{-16}~{\rm GeV}^{-1}$, assuming $g_{ap} \sim 10^{-9}$ (near the SN 1987A cooling limit~\cite{Lella:2023bfb,Fiorillo:2025gnd}).
Beyond $\sim 10$ stacked extragalactic SNe, however, the returns become marginal, as shown in the inset in Fig.~\ref{fig:fig4}. Consequently, even with a rate of one nearby SN per year, a decade-long monitoring program would be sufficient to either detect a signal or set a substantially stronger bound on the ALP couplings.

\section{Conclusions}

\label{sec:Conclusions}
Core–collapse supernovae in nearby galaxies offer a powerful and previously underutilized opportunity to explore the ALP parameter space well beyond existing limits. 
Their unique role stems from the fact that several nearby galaxies host magnetic environments—ranging from starburst-driven mG fields to cluster-scale intracluster structures—where ALP–photon conversion efficiencies exceed those of the Milky Way by 2–3 orders of magnitude.

In addition, a few core-collapse SNe are expected to occur every year  within $\sim 20 \, \mathrm{Mpc}$, providing a steady stream of potential ALP sources. Yet current gamma-ray facilities lack the full-sky coverage and time-tagging accuracy needed to exploit these events effectively. 
Our analysis shows that two realistic experimental capabilities would decisively change this situation:
\emph{(i)} a full-sky, continuously operating constellation of MeV gamma-ray detectors, and
\emph{(ii)} an external trigger capable of determining the core-collapse time to within seconds, most naturally provided by next-generation deci-hertz GW interferometers.
If these conditions are met, extragalactic SNe become a transformative ALP probe. 
A single explosion in a favorable host galaxy would surpass the SN 1987A constraint, and a decade-long program could explore $g_{ap} \times g_{a\gamma}$ at the level of $\sim 10^{-24}\,\mathrm{GeV}^{-1}$, reaching deep into currently inaccessible parameter space.

Extragalactic SN ALP searches would, thus, represent a major new frontier in \textit{multi-messenger astronomy}. 
By combining gamma rays, optical surveys, and gravitational waves, they open a realistic pathway toward \textit{extragalactic ALP astronomy}, in which the cumulative SN activity of the local Universe becomes a precision tool for fundamental physics.

\acknowledgements
\vspace{-1.pt}
We warmly thank Christopher Eckner, Damiano F.G. Fiorillo, Orion Ning, and Edoardo Vitagliano for their helpful comments on the manuscript. This article is based on work from COST Action COSMIC WISPers CA21106, supported by COST (European Cooperation in Science and Technology).
The work of AM and AL and FL was partially supported by the research grant number 2022E2J4RK "PANTHEON: Perspectives in Astroparticle and Neutrino THEory with Old and New messengers" under the program PRIN 2022 (Mission 4, Component 1,
CUP I53D23001110006) funded by the Italian Ministero dell'Universit\`a e della Ricerca (MUR) and by the European Union – Next Generation EU. AL acknowledges support from Italian MUR through the FIS 2 project FIS-2023-01577 (DD n. 23314 10-12-2024, CUP C53C24001460001), and by Istituto Nazionale di Fisica Nucleare (INFN) through the Theoretical Astroparticle Physics (TAsP) project. 
GL acknowledges support from the U.S. Department of Energy under contract number DE-AC02-76SF00515. This work is (partially) supported by ICSC – Centro Nazionale di Ricerca in High Performance Computing. 
M. Giannotti acknowledges support from the Spanish Agencia Estatal de Investigación through grant PID2019-108122GB-C31, funded by MCIN/AEI/10.13039/501100011033, and from the European Union NextGenerationEU/PRTR initiative (Planes Complementarios, Programa de Astrofísica y Física de Altas Energías). M. Giannotti also acknowledges support from grant PGC2022-126078NB-C21, `Aún más allá de los modelos estándar,'' funded by MCIN/AEI/10.13039/501100011033 and by the European Regional Development Fund (ERDF), `A way of making Europe''; from the European Union's Horizon 2020 research and innovation programme under European Research Council (ERC) grant agreement No.\ 788781 (IAXO+); and from project CNS2025-165965, ``Señales de axiones en el rango MeV desde fuentes astrofísicas hasta detectores actuales y futuros,'' funded by MICIU/AEI/10.13039/501100011033.

\nocite{*}
\bibliographystyle{bibi}
\bibliography{references}

@article{Jaeckel:2010ni,
    author = "Jaeckel, Joerg and Ringwald, Andreas",
    title = "{The Low-Energy Frontier of Particle Physics}",
    eprint = "1002.0329",
    archivePrefix = "arXiv",
    primaryClass = "hep-ph",
    reportNumber = "CPT-10-18, DESY-10-016, IPPP-10-09",
    doi = "10.1146/annurev.nucl.012809.104433",
    journal = "Ann. Rev. Nucl. Part. Sci.",
    volume = "60",
    pages = "405--437",
    year = "2010"
}

@inproceedings{Ringwald:2014vqa,
    author = "Ringwald, A.",
    title = "{Axions and Axion-Like Particles}",
    booktitle = "{49th Rencontres de Moriond on Electroweak Interactions and Unified Theories}",
    eprint = "1407.0546",
    archivePrefix = "arXiv",
    primaryClass = "hep-ph",
    reportNumber = "DESY-14-108",
    pages = "223--230",
    year = "2014"
}

@misc{AxionLimits,
  author       = {Ciaran O'Hare},
  title        = {cajohare/AxionLimits: AxionLimits},
  month        = jul,
  year         = 2020,
  publisher    = {Zenodo},
  version      = {v1.0},
  doi          = {10.5281/zenodo.3932430},
  howpublished = {\url{https://cajohare.github.io/AxionLimits/}}
}

@article{DiLuzio:2020wdo,
    author = "Di Luzio, Luca and Giannotti, Maurizio and Nardi, Enrico and Visinelli, Luca",
    title = "{The landscape of QCD axion models}",
    eprint = "2003.01100",
    archivePrefix = "arXiv",
    primaryClass = "hep-ph",
    reportNumber = "DESY 20-036, DESY-20-036",
    doi = "10.1016/j.physrep.2020.06.002",
    journal = "Phys. Rept.",
    volume = "870",
    pages = "1--117",
    year = "2020"
}

@article{IAXO:2020wwp,
    author = "Abeln, A. and others",
    collaboration = "IAXO",
    title = "{Conceptual design of BabyIAXO, the intermediate stage towards the International Axion Observatory}",
    eprint = "2010.12076",
    archivePrefix = "arXiv",
    primaryClass = "physics.ins-det",
    doi = "10.1007/JHEP05(2021)137",
    journal = "JHEP",
    volume = "05",
    pages = "137",
    year = "2021"
}

@article{IAXO:2019mpb,
    author = "Armengaud, E. and others",
    collaboration = "IAXO",
    title = "{Physics potential of the International Axion Observatory (IAXO)}",
    eprint = "1904.09155",
    archivePrefix = "arXiv",
    primaryClass = "hep-ph",
    doi = "10.1088/1475-7516/2019/06/047",
    journal = "JCAP",
    volume = "06",
    pages = "047",
    year = "2019"
}

@article{Ruz:2024gkl,
    author = "Ruz, J. and others",
    title = "{NuSTAR as an Axion Helioscope}",
    eprint = "2407.03828",
    archivePrefix = "arXiv",
    primaryClass = "astro-ph.CO",
    doi = "10.1103/18sn-hxtb",
    journal = "Phys. Rev. Lett.",
    volume = "135",
    number = "14",
    pages = "141001",
    year = "2025"
}

@misc{SNarchive,
     title = "{Garching core-collapse supernova research archive}",
     howpublished = {\url{https://wwwmpa.mpa-garching.mpg.de/ccsnarchive//}}
}

@article{DiLuzio:2021ysg,
    author = "Di Luzio, Luca and Fedele, Marco and Giannotti, Maurizio and Mescia, Federico and Nardi, Enrico",
    title = "{Stellar evolution confronts axion models}",
    eprint = "2109.10368",
    archivePrefix = "arXiv",
    primaryClass = "hep-ph",
    reportNumber = "DESY-21-141, TTP21-030, P3H-21-062",
    doi = "10.1088/1475-7516/2022/02/035",
    journal = "JCAP",
    volume = "02",
    number = "02",
    pages = "035",
    year = "2022"
}

@article{Agrawal:2021dbo,
    author = "Agrawal, Prateek and others",
    title = "{Feebly-interacting particles: FIPs 2020 workshop report}",
    eprint = "2102.12143",
    archivePrefix = "arXiv",
    primaryClass = "hep-ph",
    doi = "10.1140/epjc/s10052-021-09703-7",
    journal = "Eur. Phys. J. C",
    volume = "81",
    number = "11",
    pages = "1015",
    year = "2021"
}

@article{Giannotti:2022euq,
    author = "Giannotti, Maurizio",
    title = "{Aspects of Axions and ALPs Phenomenology}",
    eprint = "2205.06831",
    archivePrefix = "arXiv",
    primaryClass = "hep-ph",
    doi = "10.1088/1742-6596/2502/1/012003",
    journal = "J. Phys. Conf. Ser.",
    volume = "2502",
    number = "1",
    pages = "012003",
    year = "2023"
}

@article{Antel:2023hkf,
    author = "Antel, C. and others",
    title = "{Feebly-interacting particles: FIPs 2022 Workshop Report}",
    eprint = "2305.01715",
    archivePrefix = "arXiv",
    primaryClass = "hep-ph",
    reportNumber = "CERN-TH-2023-061, DESY-23-050, FERMILAB-PUB-23-149-PPD, INFN-23-14-LNF, JLAB-PHY-23-3789, LA-UR-23-21432, MITP-23-015",
    doi = "10.1140/epjc/s10052-023-12168-5",
    journal = "Eur. Phys. J. C",
    volume = "83",
    number = "12",
    pages = "1122",
    year = "2023"
}

@article{Peccei:1977hh,
    author = "Peccei, R. D. and Quinn, Helen R.",
    title = "{CP Conservation in the Presence of Instantons}",
    reportNumber = "ITP-568-STANFORD",
    doi = "10.1103/PhysRevLett.38.1440",
    journal = "Phys. Rev. Lett.",
    volume = "38",
    pages = "1440--1443",
    year = "1977"
}

@article{Peccei:1977ur,
    author = "Peccei, R. D. and Quinn, Helen R.",
    title = "{Constraints Imposed by CP Conservation in the Presence of Instantons}",
    reportNumber = "ITP-572-STANFORD",
    doi = "10.1103/PhysRevD.16.1791",
    journal = "Phys. Rev. D",
    volume = "16",
    pages = "1791--1797",
    year = "1977"
}

@article{Weinberg:1977ma,
    author = "Weinberg, Steven",
    title = "{A New Light Boson?}",
    reportNumber = "HUTP-77/A074",
    doi = "10.1103/PhysRevLett.40.223",
    journal = "Phys. Rev. Lett.",
    volume = "40",
    pages = "223--226",
    year = "1978"
}

@article{Wilczek:1977pj,
    author = "Wilczek, Frank",
    title = "{Problem of Strong  $P$  and  $T$  Invariance in the Presence of Instantons}",
    reportNumber = "Print-77-0939 (COLUMBIA)",
    doi = "10.1103/PhysRevLett.40.279",
    journal = "Phys. Rev. Lett.",
    volume = "40",
    pages = "279--282",
    year = "1978"
}

@article{Arvanitaki:2009fg,
    author = "Arvanitaki, Asimina and Dimopoulos, Savas and Dubovsky, Sergei and Kaloper, Nemanja and March-Russell, John",
    title = "{String Axiverse}",
    eprint = "0905.4720",
    archivePrefix = "arXiv",
    primaryClass = "hep-th",
    doi = "10.1103/PhysRevD.81.123530",
    journal = "Phys. Rev. D",
    volume = "81",
    pages = "123530",
    year = "2010"
}

@article{Cicoli:2012sz,
    author = "Cicoli, Michele and Goodsell, Mark and Ringwald, Andreas",
    title = "{The type IIB string axiverse and its low-energy phenomenology}",
    eprint = "1206.0819",
    archivePrefix = "arXiv",
    primaryClass = "hep-th",
    reportNumber = "DESY-12-058, CERN-PH-TH-2012-153",
    doi = "10.1007/JHEP10(2012)146",
    journal = "JHEP",
    volume = "10",
    pages = "146",
    year = "2012"
}

@article{Cicoli:2023opf,
    author = "Cicoli, Michele and Conlon, Joseph P. and Maharana, Anshuman and Parameswaran, Susha and Quevedo, Fernando and Zavala, Ivonne",
    title = "{String cosmology: From the early universe to today}",
    eprint = "2303.04819",
    archivePrefix = "arXiv",
    primaryClass = "hep-th",
    doi = "10.1016/j.physrep.2024.01.002",
    journal = "Phys. Rept.",
    volume = "1059",
    pages = "1--155",
    year = "2024"
}

@article{Abbott:1982af,
    author = "Abbott, L. F. and Sikivie, P.",
    editor = "Srednicki, M. A.",
    title = "{A Cosmological Bound on the Invisible Axion}",
    reportNumber = "PRINT-82-0695 (BRANDEIS)",
    doi = "10.1016/0370-2693(83)90638-X",
    journal = "Phys. Lett. B",
    volume = "120",
    pages = "133--136",
    year = "1983"
}

@article{Dine:1982ah,
    author = "Dine, Michael and Fischler, Willy",
    editor = "Srednicki, M. A.",
    title = "{The Not So Harmless Axion}",
    reportNumber = "UPR-0201T",
    doi = "10.1016/0370-2693(83)90639-1",
    journal = "Phys. Lett. B",
    volume = "120",
    pages = "137--141",
    year = "1983"
}

@article{Preskill:1982cy,
    author = "Preskill, John and Wise, Mark B. and Wilczek, Frank",
    editor = "Srednicki, M. A.",
    title = "{Cosmology of the Invisible Axion}",
    reportNumber = "HUTP-82-A048, NSF-ITP-82-103",
    doi = "10.1016/0370-2693(83)90637-8",
    journal = "Phys. Lett. B",
    volume = "120",
    pages = "127--132",
    year = "1983"
}

@article{Arias:2012az,
    author = "Arias, Paola and Cadamuro, Davide and Goodsell, Mark and Jaeckel, Joerg and Redondo, Javier and Ringwald, Andreas",
    title = "{WISPy Cold Dark Matter}",
    eprint = "1201.5902",
    archivePrefix = "arXiv",
    primaryClass = "hep-ph",
    reportNumber = "DESY-11-226, MPP-2011-140, CERN-PH-TH-2011-323, IPPP-11-80, DCPT-11-160",
    doi = "10.1088/1475-7516/2012/06/013",
    journal = "JCAP",
    volume = "06",
    pages = "013",
    year = "2012"
}

@inproceedings{Adams:2022pbo,
    author = "Adams, C. B. and others",
    title = "{Axion Dark Matter}",
    booktitle = "{Snowmass 2021}",
    eprint = "2203.14923",
    archivePrefix = "arXiv",
    primaryClass = "hep-ex",
    reportNumber = "FERMILAB-CONF-22-996-PPD-T",
    month = "3",
    year = "2022"
}

@article{Giannotti:2015kwo,
    author = "Giannotti, Maurizio and Irastorza, Igor and Redondo, Javier and Ringwald, Andreas",
    title = "{Cool WISPs for stellar cooling excesses}",
    eprint = "1512.08108",
    archivePrefix = "arXiv",
    primaryClass = "astro-ph.HE",
    reportNumber = "DESY-15-245",
    doi = "10.1088/1475-7516/2016/05/057",
    journal = "JCAP",
    volume = "05",
    pages = "057",
    year = "2016"
}

@article{Giannotti:2017hny,
    author = "Giannotti, Maurizio and Irastorza, Igor G. and Redondo, Javier and Ringwald, Andreas and Saikawa, Ken'ichi",
    title = "{Stellar Recipes for Axion Hunters}",
    eprint = "1708.02111",
    archivePrefix = "arXiv",
    primaryClass = "hep-ph",
    reportNumber = "DESY-17-116",
    doi = "10.1088/1475-7516/2017/10/010",
    journal = "JCAP",
    volume = "10",
    pages = "010",
    year = "2017"
}

@article{Galanti:2022chk,
    author = "Galanti, Giorgio and Nava, Lara and Roncadelli, Marco and Tavecchio, Fabrizio and Bonnoli, Giacomo",
    title = "{Observability of the Very-High-Energy Emission from GRB 221009A}",
    eprint = "2210.05659",
    archivePrefix = "arXiv",
    primaryClass = "astro-ph.HE",
    doi = "10.1103/PhysRevLett.131.251001",
    journal = "Phys. Rev. Lett.",
    volume = "131",
    number = "25",
    pages = "251001",
    year = "2023"
}

@book{Raffelt:1996wa,
    author = "Raffelt, G. G.",
    title = "{Stars as laboratories for fundamental physics}: {The astrophysics of neutrinos, axions, and other weakly interacting particles}",
    publisher = "University of Chicago Press",
    isbn = "978-0-226-70272-8",
    month = "5",
    year = "1996"
}

@article{Raffelt:1999tx,
    author = "Raffelt, Georg G.",
    title = "{Particle physics from stars}",
    eprint = "hep-ph/9903472",
    archivePrefix = "arXiv",
    doi = "10.1146/annurev.nucl.49.1.163",
    journal = "Ann. Rev. Nucl. Part. Sci.",
    volume = "49",
    pages = "163--216",
    year = "1999"
}

@article{Caputo:2024oqc,
    author = "Caputo, Andrea and Raffelt, Georg",
    title = "{Astrophysical Axion Bounds: The 2024 Edition}",
    eprint = "2401.13728",
    archivePrefix = "arXiv",
    primaryClass = "hep-ph",
    reportNumber = "MPP-2024-13, CERN-TH-2024-013",
    doi = "10.22323/1.454.0041",
    journal = "PoS",
    volume = "COSMICWISPers",
    pages = "041",
    year = "2024"
}

@article{Carenza:2024ehj,
    author = "Carenza, Pierluca and Giannotti, Maurizio and Isern, Jordi and Mirizzi, Alessandro and Straniero, Oscar",
    title = "{Axion astrophysics}",
    eprint = "2411.02492",
    archivePrefix = "arXiv",
    primaryClass = "hep-ph",
    reportNumber = "BARI-TH/66-24",
    doi = "10.1016/j.physrep.2025.02.002",
    journal = "Phys. Rept.",
    volume = "1117",
    pages = "1--102",
    year = "2025"
}

@article{vanBibber:1988ge,
    author = "van Bibber, K. and McIntyre, P. M. and Morris, D. E. and Raffelt, G. G.",
    title = "{A Practical Laboratory Detector for Solar Axions}",
    reportNumber = "LBL-25908",
    doi = "10.1103/PhysRevD.39.2089",
    journal = "Phys. Rev. D",
    volume = "39",
    pages = "2089",
    year = "1989"
}

@article{Kawamura:2020pcg,
    author = "Kawamura, Seiji and others",
    title = "{Current status of space gravitational wave antenna DECIGO and B-DECIGO}",
    eprint = "2006.13545",
    archivePrefix = "arXiv",
    primaryClass = "gr-qc",
    doi = "10.1093/ptep/ptab019",
    journal = "PTEP",
    volume = "2021",
    number = "5",
    pages = "05A105",
    year = "2021"
}

@article{Dicus:1978fp,
    author = "Dicus, Duane A. and Kolb, Edward W. and Teplitz, Vigdor L. and Wagoner, Robert V.",
    title = "{Astrophysical Bounds on the Masses of Axions and Higgs Particles}",
    reportNumber = "Print-78-0407 (TEXAS), ORO-3992-337",
    doi = "10.1103/PhysRevD.18.1829",
    journal = "Phys. Rev. D",
    volume = "18",
    pages = "1829",
    year = "1978"
}

@article{Raffelt:1985nk,
    author = "Raffelt, Georg G.",
    title = "{Astrophysical axion bounds diminished by screening effects}",
    reportNumber = "MPI-PAE-PTH-51-85",
    doi = "10.1103/PhysRevD.33.897",
    journal = "Phys. Rev. D",
    volume = "33",
    pages = "897",
    year = "1986"
}

@article{Ning:2025tit,
    author = "Ning, Orion and Safdi, Benjamin R.",
    title = "{Probing the Axion-Electron Coupling with NuSTAR Observations of Galaxies}",
    eprint = "2503.09682",
    archivePrefix = "arXiv",
    primaryClass = "hep-ph",
    month = "3",
    year = "2025"
}

@article{Kamiokande-II:1987idp,
    author = "Hirata, K. and others",
    editor = "Wali, K. C.",
    collaboration = "Kamiokande-II",
    title = "{Observation of a Neutrino Burst from the Supernova SN 1987a}",
    reportNumber = "UT-ICEPP-87-01, UPR-142E",
    doi = "10.1103/PhysRevLett.58.1490",
    journal = "Phys. Rev. Lett.",
    volume = "58",
    pages = "1490--1493",
    year = "1987"
}

@article{Hirata:1988ad,
    author = "Hirata, K. S. and others",
    title = "{Observation in the Kamiokande-II Detector of the Neutrino Burst from Supernova SN 1987a}",
    doi = "10.1103/PhysRevD.38.448",
    journal = "Phys. Rev. D",
    volume = "38",
    pages = "448--458",
    year = "1988"
}

@article{Bionta:1987qt,
    author = "Bionta, R. M. and others",
    title = "{Observation of a Neutrino Burst in Coincidence with Supernova SN 1987a in the Large Magellanic Cloud}",
    reportNumber = "UCI-NEUTRINO-87-10",
    doi = "10.1103/PhysRevLett.58.1494",
    journal = "Phys. Rev. Lett.",
    volume = "58",
    pages = "1494",
    year = "1987"
}

@article{IMB:1988suc,
    author = "Bratton, C. B. and others",
    collaboration = "IMB",
    title = "{Angular Distribution of Events From Sn1987a}",
    reportNumber = "UM-PDK-88-1",
    doi = "10.1103/PhysRevD.37.3361",
    journal = "Phys. Rev. D",
    volume = "37",
    pages = "3361",
    year = "1988"
}

@article{Alekseev:1988gp,
    author = "Alekseev, E. N. and Alekseeva, L. N. and Krivosheina, I. V. and Volchenko, V. I.",
    title = "{Detection of the Neutrino Signal From {SN1987A} in the {LMC} Using the Inr Baksan Underground Scintillation Telescope}",
    doi = "10.1016/0370-2693(88)91651-6",
    journal = "Phys. Lett. B",
    volume = "205",
    pages = "209--214",
    year = "1988"
}

@article{Grifols:1996id,
    author = "Grifols, J. A. and Masso, E. and Toldra, R.",
    title = "{Gamma-rays from SN1987A due to pseudoscalar conversion}",
    eprint = "astro-ph/9606028",
    archivePrefix = "arXiv",
    reportNumber = "UAB-FT-391",
    doi = "10.1103/PhysRevLett.77.2372",
    journal = "Phys. Rev. Lett.",
    volume = "77",
    pages = "2372--2375",
    year = "1996"
}

@article{Brockway:1996yr,
    author = "Brockway, Jack W. and Carlson, Eric D. and Raffelt, Georg G.",
    title = "{SN1987A gamma-ray limits on the conversion of pseudoscalars}",
    eprint = "astro-ph/9605197",
    archivePrefix = "arXiv",
    reportNumber = "MPI-PHT-96-42",
    doi = "10.1016/0370-2693(96)00778-2",
    journal = "Phys. Lett. B",
    volume = "383",
    pages = "439--443",
    year = "1996"
}

@article{Mukhopadhyay:2021gox,
    author = "Mukhopadhyay, Mainak and Lin, Zidu and Lunardini, Cecilia",
    title = "{Memory-triggered supernova neutrino detection}",
    eprint = "2110.14657",
    archivePrefix = "arXiv",
    primaryClass = "astro-ph.HE",
    doi = "10.1103/PhysRevD.106.043020",
    journal = "Phys. Rev. D",
    volume = "106",
    number = "4",
    pages = "043020",
    year = "2022"
}

@article{Carenza:2019pxu,
    author = "Carenza, Pierluca and Fischer, Tobias and Giannotti, Maurizio and Guo, Gang and Mart{\'\i}nez-Pinedo, Gabriel and Mirizzi, Alessandro",
    title = "{Improved axion emissivity from a supernova via nucleon-nucleon bremsstrahlung}",
    eprint = "1906.11844",
    archivePrefix = "arXiv",
    primaryClass = "hep-ph",
    doi = "10.1088/1475-7516/2019/10/016",
    journal = "JCAP",
    volume = "10",
    number = "10",
    pages = "016",
    year = "2019",
    note = "[Erratum: JCAP 05, E01 (2020)]"
}

@article{Brinkmann:1988vi,
    author = "Brinkmann, Ralf Peter and Turner, Michael S.",
    title = "{Numerical Rates for Nucleon-Nucleon Axion Bremsstrahlung}",
    reportNumber = "FERMILAB-PUB-88-029-A",
    doi = "10.1103/PhysRevD.38.2338",
    journal = "Phys. Rev. D",
    volume = "38",
    pages = "2338",
    year = "1988"
}

@article{Burrows:1988ah,
    author = "Burrows, Adam and Turner, Michael S. and Brinkmann, R. P.",
    title = "{Axions and SN 1987a}",
    reportNumber = "FERMILAB-PUB-88-105-A",
    doi = "10.1103/PhysRevD.39.1020",
    journal = "Phys. Rev. D",
    volume = "39",
    pages = "1020",
    year = "1989"
}

@article{Candon:2024eah,
    author = "Cand{\'o}n, Francisco R. and Fiorillo, Damiano F. G. and Lucente, Giuseppe and Vitagliano, Edoardo and Vogel, Julia K.",
    title = "{NuSTAR Bounds on Radiatively Decaying Particles from M82}",
    eprint = "2412.03660",
    archivePrefix = "arXiv",
    primaryClass = "hep-ph",
    doi = "10.1103/PhysRevLett.134.171004",
    journal = "Phys. Rev. Lett.",
    volume = "134",
    number = "17",
    pages = "171004",
    year = "2025"
}

@article{Meyer:2016wrm,
    author = "Meyer, M. and Giannotti, M. and Mirizzi, A. and Conrad, J. and S{\'a}nchez-Conde, M. A.",
    title = "{Fermi Large Area Telescope as a Galactic Supernovae Axionscope}",
    eprint = "1609.02350",
    archivePrefix = "arXiv",
    primaryClass = "astro-ph.HE",
    doi = "10.1103/PhysRevLett.118.011103",
    journal = "Phys. Rev. Lett.",
    volume = "118",
    number = "1",
    pages = "011103",
    year = "2017"
}

@article{Keil:1996ju,
    author = "Keil, Wolfgang and Janka, Hans-Thomas and Schramm, David N. and Sigl, Gunter and Turner, Michael S. and Ellis, John R.",
    title = "{A Fresh look at axions and SN-1987A}",
    eprint = "astro-ph/9612222",
    archivePrefix = "arXiv",
    reportNumber = "FERMILAB-PUB-95-406-A, CERN-TH-96-112",
    doi = "10.1103/PhysRevD.56.2419",
    journal = "Phys. Rev. D",
    volume = "56",
    pages = "2419--2432",
    year = "1997"
}

@article{Carenza:2020cis,
    author = "Carenza, Pierluca and Fore, Bryce and Giannotti, Maurizio and Mirizzi, Alessandro and Reddy, Sanjay",
    title = "{Enhanced Supernova Axion Emission and its Implications}",
    eprint = "2010.02943",
    archivePrefix = "arXiv",
    primaryClass = "hep-ph",
    reportNumber = "INT-PUB-20-039",
    doi = "10.1103/PhysRevLett.126.071102",
    journal = "Phys. Rev. Lett.",
    volume = "126",
    number = "7",
    pages = "071102",
    year = "2021"
}

@article{Payez:2014xsa,
    author = "Payez, Alexandre and Evoli, Carmelo and Fischer, Tobias and Giannotti, Maurizio and Mirizzi, Alessandro and Ringwald, Andreas",
    title = "{Revisiting the SN1987A gamma-ray limit on ultralight axion-like particles}",
    eprint = "1410.3747",
    archivePrefix = "arXiv",
    primaryClass = "astro-ph.HE",
    reportNumber = "DESY-14-164",
    doi = "10.1088/1475-7516/2015/02/006",
    journal = "JCAP",
    volume = "02",
    pages = "006",
    year = "2015"
}

@article{Hoof:2022xbe,
    author = "Hoof, Sebastian and Schulz, Lena",
    title = "{Updated constraints on axion-like particles from temporal information in supernova SN1987A gamma-ray data}",
    eprint = "2212.09764",
    archivePrefix = "arXiv",
    primaryClass = "hep-ph",
    reportNumber = "TTP22-072",
    doi = "10.1088/1475-7516/2023/03/054",
    journal = "JCAP",
    volume = "03",
    pages = "054",
    year = "2023"
}

@article{Manzari:2024jns,
    author = "Manzari, Claudio Andrea and Park, Yujin and Safdi, Benjamin R. and Savoray, Inbar",
    title = "{Supernova Axions Convert to Gamma Rays in Magnetic Fields of Progenitor Stars}",
    eprint = "2405.19393",
    archivePrefix = "arXiv",
    primaryClass = "hep-ph",
    doi = "10.1103/PhysRevLett.133.211002",
    journal = "Phys. Rev. Lett.",
    volume = "133",
    number = "21",
    pages = "211002",
    year = "2024"
}

@article{Fiorillo:2025gnd,
    author = "Fiorillo, Damiano F. G. and Gil Muyor, {\'A}ngel and Janka, Hans-Thomas and Raffelt, Georg G. and Vitagliano, Edoardo",
    title = "{Axion-photon conversion in transient compact stars: Systematics, constraints, and opportunities}",
    eprint = "2509.13322"
}

@article{Rozwadowska:2020nab,
    author = "Rozwadowska, Karolina and Vissani, Francesco and Cappellaro, Enrico",
    title = "{On the rate of core collapse supernovae in the milky way}",
    eprint = "2009.03438",
    archivePrefix = "arXiv",
    primaryClass = "astro-ph.HE",
    doi = "10.1016/j.newast.2020.101498",
    journal = "New Astron.",
    volume = "83",
    pages = "101498",
    year = "2021"
}

@article{Meyer:2020vzy,
    author = "Meyer, Manuel and Petrushevska, Tanja",
    title = "{Search for Axionlike-Particle-Induced Prompt $\gamma$-Ray Emission from Extragalactic Core-Collapse Supernovae with the $Fermi$ Large Area Telescope}",
    eprint = "2006.06722",
    archivePrefix = "arXiv",
    primaryClass = "astro-ph.HE",
    doi = "10.1103/PhysRevLett.124.231101",
    journal = "Phys. Rev. Lett.",
    volume = "124",
    number = "23",
    pages = "231101",
    year = "2020",
    note = "[Erratum: Phys.Rev.Lett. 125, 119901 (2020)]"
}

@article{Mukhopadhyay:2021zbt,
    author = "Mukhopadhyay, Mainak and Cardona, Carlos and Lunardini, Cecilia",
    title = "{The neutrino gravitational memory from a core collapse supernova: phenomenology and physics potential}",
    eprint = "2105.05862",
    archivePrefix = "arXiv",
    primaryClass = "astro-ph.HE",
    doi = "10.1088/1475-7516/2021/07/055",
    journal = "JCAP",
    volume = "07",
    pages = "055",
    year = "2021"
}

@article{Seto:2001qf,
    author = "Seto, Naoki and Kawamura, Seiji and Nakamura, Takashi",
    title = "{Possibility of direct measurement of the acceleration of the universe using 0.1-Hz band laser interferometer gravitational wave antenna in space}",
    eprint = "astro-ph/0108011",
    archivePrefix = "arXiv",
    doi = "10.1103/PhysRevLett.87.221103",
    journal = "Phys. Rev. Lett.",
    volume = "87",
    pages = "221103",
    year = "2001"
}

@article{Yagi:2011wg,
    author = "Yagi, Kent and Seto, Naoki",
    title = "{Detector configuration of DECIGO/BBO and identification of cosmological neutron-star binaries}",
    eprint = "1101.3940",
    archivePrefix = "arXiv",
    primaryClass = "astro-ph.CO",
    doi = "10.1103/PhysRevD.83.044011",
    journal = "Phys. Rev. D",
    volume = "83",
    pages = "044011",
    year = "2011",
    note = "[Erratum: Phys.Rev.D 95, 109901 (2017)]"
}

@article{Raffelt:1993ix,
    author = "Raffelt, Georg and Seckel, David",
    title = "{A selfconsistent approach to neutral current processes in supernova cores}",
    eprint = "astro-ph/9312019",
    archivePrefix = "arXiv",
    reportNumber = "MPI-PH-93-90, BA-93-43",
    doi = "10.1103/PhysRevD.52.1780",
    journal = "Phys. Rev. D",
    volume = "52",
    pages = "1780--1799",
    year = "1995"
}

@article{Carena:1988kr,
    author = "Carena, Marcela and Peccei, R. D.",
    title = "{The Effective Lagrangian for Axion Emission From {SN1987A}}",
    reportNumber = "DESY-88-182",
    doi = "10.1103/PhysRevD.40.652",
    journal = "Phys. Rev. D",
    volume = "40",
    pages = "652",
    year = "1989"
}

@article{Lella:2022uwi,
    author = "Lella, Alessandro and Carenza, Pierluca and Lucente, Giuseppe and Giannotti, Maurizio and Mirizzi, Alessandro",
    title = "{Protoneutron stars as cosmic factories for massive axionlike particles}",
    eprint = "2211.13760",
    archivePrefix = "arXiv",
    primaryClass = "hep-ph",
    doi = "10.1103/PhysRevD.107.103017",
    journal = "Phys. Rev. D",
    volume = "107",
    number = "10",
    pages = "103017",
    year = "2023"
}

@article{Lella:2023bfb,
    author = "Lella, Alessandro and Carenza, Pierluca and Co', Giampaolo and Lucente, Giuseppe and Giannotti, Maurizio and Mirizzi, Alessandro and Rauscher, Thomas",
    title = "{Getting the most on supernova axions}",
    eprint = "2306.01048",
    archivePrefix = "arXiv",
    primaryClass = "hep-ph",
    doi = "10.1103/PhysRevD.109.023001",
    journal = "Phys. Rev. D",
    volume = "109",
    number = "2",
    pages = "023001",
    year = "2024"
}

@article{Rampp:2002bq,
    author = "Rampp, Markus and Janka, H. Thomas",
    title = "{Radiation hydrodynamics with neutrinos: Variable Eddington factor method for core collapse supernova simulations}",
    eprint = "astro-ph/0203101",
    archivePrefix = "arXiv",
    reportNumber = "MPA1400E",
    doi = "10.1051/0004-6361:20021398",
    journal = "Astron. Astrophys.",
    volume = "396",
    pages = "361",
    year = "2002"
}

@article{Hempel:2009mc,
    author = "Hempel, Matthias and Schaffner-Bielich, Jurgen",
    title = "{Statistical Model for a Complete Supernova Equation of State}",
    eprint = "0911.4073",
    archivePrefix = "arXiv",
    primaryClass = "nucl-th",
    doi = "10.1016/j.nuclphysa.2010.02.010",
    journal = "Nucl. Phys. A",
    volume = "837",
    pages = "210--254",
    year = "2010"
}

@article{Steiner:2012rk,
    author = "Steiner, Andrew W. and Hempel, Matthias and Fischer, Tobias",
    title = "{Core-collapse supernova equations of state based on neutron star observations}",
    eprint = "1207.2184",
    archivePrefix = "arXiv",
    primaryClass = "astro-ph.SR",
    reportNumber = "INT-PUB-12-033",
    doi = "10.1088/0004-637X/774/1/17",
    journal = "Astrophys. J.",
    volume = "774",
    pages = "17",
    year = "2013"
}

@article{Sukhbold:2017cnt,
    author = "Sukhbold, Tuguldur and Woosley, Stan and Heger, Alexander",
    title = "{A High-resolution Study of Presupernova Core Structure}",
    eprint = "1710.03243",
    archivePrefix = "arXiv",
    primaryClass = "astro-ph.HE",
    doi = "10.3847/1538-4357/aac2da",
    journal = "Astrophys. J.",
    volume = "860",
    number = "2",
    pages = "93",
    year = "2018"
}

@article{Fore:2019wib,
    author = "Fore, Bryce and Reddy, Sanjay",
    title = "{Pions in hot dense matter and their astrophysical implications}",
    eprint = "1911.02632",
    archivePrefix = "arXiv",
    primaryClass = "astro-ph.HE",
    reportNumber = "INT-PUB-19-046",
    doi = "10.1103/PhysRevC.101.035809",
    journal = "Phys. Rev. C",
    volume = "101",
    number = "3",
    pages = "035809",
    year = "2020"
}

@article{Raffelt:1987yu,
    author = "Raffelt, Georg G. and Dearborn, David S. P.",
    title = "{Bounds on Hadronic Axions From Stellar Evolution}",
    reportNumber = "UCRL-96161-REV, UCRL-96161",
    doi = "10.1103/PhysRevD.36.2211",
    journal = "Phys. Rev. D",
    volume = "36",
    pages = "2211",
    year = "1987"
}

@article{Raffelt:1987im,
    author = "Raffelt, Georg and Stodolsky, Leo",
    title = "{Mixing of the Photon with Low Mass Particles}",
    reportNumber = "MPI-PAE/PTh-54/87",
    doi = "10.1103/PhysRevD.37.1237",
    journal = "Phys. Rev. D",
    volume = "37",
    pages = "1237",
    year = "1988"
}

@article{Ning:2024eky,
    author = "Ning, Orion and Safdi, Benjamin R.",
    title = "{Leading Axion-Photon Sensitivity with NuSTAR Observations of M82 and M87}",
    eprint = "2404.14476",
    archivePrefix = "arXiv",
    primaryClass = "hep-ph",
    doi = "10.1103/PhysRevLett.134.171003",
    journal = "Phys. Rev. Lett.",
    volume = "134",
    number = "17",
    pages = "171003",
    year = "2025"
}

@article{Lopez_Rodriguez_2021,
   title={The Strength and Structure of the Magnetic Field in the Galactic Outflow of Messier 82},
   volume={914},
   ISSN={1538-4357},
   url={http://dx.doi.org/10.3847/1538-4357/abf934},
   DOI={10.3847/1538-4357/abf934},
   number={1},
   journal={The Astrophysical Journal},
   publisher={American Astronomical Society},
   author={Lopez-Rodriguez, Enrique and Guerra, Jordan A. and Asgari-Targhi, Mahboubeh and Schmelz, Joan T.},
   year={2021},
   month=jun, pages={24} }

@article{Nelson:2019jkf,
    author = "Nelson, Dylan and Pillepich, Annalisa and Springel, Volker and Pakmor, Ruediger and Weinberger, Rainer and Genel, Shy and Torrey, Paul and Vogelsberger, Mark and Marinacci, Federico and Hernquist, Lars",
    title = "{First Results from the TNG50 Simulation: Galactic outflows driven by supernovae and black hole feedback}",
    eprint = "1902.05554",
    archivePrefix = "arXiv",
    primaryClass = "astro-ph.GA",
    doi = "10.1093/mnras/stz2306",
    journal = "Mon. Not. Roy. Astron. Soc.",
    volume = "490",
    number = "3",
    pages = "3234--3261",
    year = "2019"
}

@article{Marinacci:2017wew,
    author = "Marinacci, Federico and others",
    title = "{First results from the IllustrisTNG simulations: radio haloes and magnetic fields}",
    eprint = "1707.03396",
    archivePrefix = "arXiv",
    primaryClass = "astro-ph.CO",
    doi = "10.1093/mnras/sty2206",
    journal = "Mon. Not. Roy. Astron. Soc.",
    volume = "480",
    number = "4",
    pages = "5113--5139",
    year = "2018"
}

@article{Horns:2012kw,
    author = "Horns, Dieter and Maccione, Luca and Meyer, Manuel and Mirizzi, Alessandro and Montanino, Daniele and Roncadelli, Marco",
    title = "{Hardening of TeV gamma spectrum of AGNs in galaxy clusters by conversions of photons into axion-like particles}",
    eprint = "1207.0776",
    archivePrefix = "arXiv",
    primaryClass = "astro-ph.HE",
    reportNumber = "LMU-ASC-44-12, MPP-2012-108",
    doi = "10.1103/PhysRevD.86.075024",
    journal = "Phys. Rev. D",
    volume = "86",
    pages = "075024",
    year = "2012"
}

@article{Calore:2023srn,
    author = "Calore, Francesca and Carenza, Pierluca and Eckner, Christopher and Giannotti, Maurizio and Lucente, Giuseppe and Mirizzi, Alessandro and Sivo, Francesco",
    title = "{Uncovering axionlike particles in supernova gamma-ray spectra}",
    eprint = "2306.03925",
    archivePrefix = "arXiv",
    primaryClass = "astro-ph.HE",
    reportNumber = "LAPTH-031/23",
    doi = "10.1103/PhysRevD.109.043010",
    journal = "Phys. Rev. D",
    volume = "109",
    number = "4",
    pages = "043010",
    year = "2024"
}

@article{Heston:2023gbx,
    author = "Heston, Sean and Kehoe, Emily and Suwa, Yudai and Horiuchi, Shunsaku",
    title = "{Timing coincidence search for supernova neutrinos with optical transient surveys}",
    eprint = "2302.04884",
    archivePrefix = "arXiv",
    primaryClass = "astro-ph.HE",
    doi = "10.1103/PhysRevD.107.123034",
    journal = "Phys. Rev. D",
    volume = "107",
    number = "12",
    pages = "123034",
    year = "2023"
}

@article{Tartaglia:2017yym,
    author = "Tartaglia, L. and others",
    title = "{The early detection and follow-up of the highly obscured Type II supernova 2016ija/DLT16am}",
    eprint = "1711.03940",
    archivePrefix = "arXiv",
    primaryClass = "astro-ph.HE",
    doi = "10.3847/1538-4357/aaa014",
    journal = "Astrophys. J.",
    volume = "853",
    number = "1",
    pages = "62",
    year = "2018"
}

@article{Cowen_2010,
   title={Estimating the explosion time of core-collapse supernovae from their optical light curves},
   volume={33},
   ISSN={0927-6505},
   url={http://dx.doi.org/10.1016/j.astropartphys.2009.10.007},
   DOI={10.1016/j.astropartphys.2009.10.007},
   number={1},
   journal={Astroparticle Physics},
   publisher={Elsevier BV},
   author={Cowen, D.F. and Franckowiak, A. and Kowalski, M.},
   year={2010},
   month=feb, pages={19–23} }

@article{Ning:2025kyu,
    author = "Ning, Orion and Ray, Anupam and Safdi, Benjamin R.",
    title = "{Axion lines from nuclear de-excitations in galactic stellar populations}",
    eprint = "2509.03569",
    archivePrefix = "arXiv",
    primaryClass = "hep-ph",
    reportNumber = "CERN-TH-2025-162, N3AS-25-014",
    month = "9",
    year = "2025"
}

@article{Ando:2005ka,
    author = "Ando, Shin'ichiro and Beacom, John F. and Yuksel, Hasan",
    title = "{Detection of neutrinos from supernovae in nearby galaxies}",
    eprint = "astro-ph/0503321",
    archivePrefix = "arXiv",
    reportNumber = "UTAP-521",
    doi = "10.1103/PhysRevLett.95.171101",
    journal = "Phys. Rev. Lett.",
    volume = "95",
    pages = "171101",
    year = "2005"
}

@INPROCEEDINGS{2021AAS...23742305G,
       author = {{Gal-Yam}, A.},
        title = "{The TNS alert system}",
    booktitle = {American Astronomical Society Meeting Abstracts \#237},
         year = 2021,
       series = {American Astronomical Society Meeting Abstracts},
       volume = {237},
        month = jan,
          eid = {423.05},
        pages = {423.05},
       adsurl = {https://ui.adsabs.harvard.edu/abs/2021AAS...23742305G}
}

@article{Khademi_2023,
   title={Influence of Magnetic Fields on the Gas Rotation in the Galaxy NGC 6946},
   volume={945},
   ISSN={1538-4357},
   url={http://dx.doi.org/10.3847/1538-4357/acb99b},
   DOI={10.3847/1538-4357/acb99b},
   number={1},
   journal={The Astrophysical Journal},
   publisher={American Astronomical Society},
   author={Khademi, M. and Nasiri, S. and Tabatabaei, F. S.},
   year={2023},
   month=mar, pages={36} }

@article{Nakamura:2016thx,
    author = "Nakamura, Ko and Horiuchi, Shunsaku and Tanaka, Masaomi and Hayama, Kazuhiro and Takiwaki, Tomoya and Kotake, Kei",
    title = "{Multi-messenger signals from core-collapse supernovae}",
    doi = "10.1017/S1743921317001193",
    journal = "IAU Symp.",
    volume = "329",
    pages = "428",
    year = "2016"
}

@article{Kennicutt:2008ce,
    author = "Kennicutt, Jr., Robert C. and Lee, Janice C. and Funes S. J., Jose G. and Sakai, Shoko and Akiyama, Sanae",
    title = "{An H-alpha Imaging Survey of Galaxies in the Local 11 Mpc Volume}",
    eprint = "0807.2035",
    archivePrefix = "arXiv",
    primaryClass = "astro-ph",
    doi = "10.1086/590058",
    journal = "Astrophys. J. Suppl.",
    volume = "178",
    pages = "247",
    year = "2008"
}

@article{Tully:2013wqa,
    author = "Tully, R. Brent and others",
    title = "{Cosmicflows-2: The Data}",
    eprint = "1307.7213",
    archivePrefix = "arXiv",
    primaryClass = "astro-ph.CO",
    doi = "10.1088/0004-6256/146/4/86",
    journal = "Astron. J.",
    volume = "146",
    pages = "86",
    year = "2013"
}

@article{Ochsenbein:2000th,
    author = "Ochsenbein, Francois and Bauer, Patricia and Marcout, James",
    title = "{The VizieR database of astronomical catalogues}",
    eprint = "astro-ph/0002122",
    archivePrefix = "arXiv",
    doi = "10.1051/aas:2000169",
    journal = "Astron. Astrophys. Suppl. Ser.",
    volume = "143",
    pages = "23--32",
    year = "2000"
}

@ARTICLE{2008CBET.1381....1W,
       author = {{Arbour}, R. and {Boles}, T.},
        title = "{Supernova 2008S in NGC 6946}",
      journal = {Central Bureau Electronic Telegrams},
         year = 2008,
        month = feb,
       volume = {1234},
        pages = {1},
       adsurl = {https://ui.adsabs.harvard.edu/abs/2008CBET.1234....1A}
}

@ARTICLE{2008CBET.1280....1M,
       author = {{Mostardi}, R. and {Li}, W. and {Filippenko}, A.~V.},
        title = "{Possible Supernova in NGC 4490}",
      journal = {Central Bureau Electronic Telegrams},
         year = 2008,
        month = mar,
       volume = {1280},
        pages = {1},
       adsurl = {https://ui.adsabs.harvard.edu/abs/2008CBET.1280....1M}
}

@ARTICLE{2008ATel.1452....1S,
       author = {{Monard}, L.~A.~G.},
        title = "{Supernova 2008bk in NGC 7793}",
      journal = {Central Bureau Electronic Telegrams},
         year = 2008,
        month = mar,
       volume = {1315},
        pages = {1},
       adsurl = {https://ui.adsabs.harvard.edu/abs/2008CBET.1315....1M}
}

@article{Brunthaler:2010bm,
    author = "Brunthaler, A. and others",
    title = "{VLBI observations of SN 2008iz: I. Expansion velocity and limits on anisotropic expansion}",
    eprint = "1003.4665",
    archivePrefix = "arXiv",
    primaryClass = "astro-ph.CO",
    doi = "10.1051/0004-6361/201014133",
    journal = "Astron. Astrophys.",
    volume = "516",
    pages = "A27",
    year = "2010"
}

@article{Prieto:2011tb,
    author = "Prieto, Jose L. and others",
    title = "{SN 2008jb: A 'Lost' Core-Collapse Supernova in a Star-Forming Dwarf Galaxy at {\textasciitilde}10 Mpc}",
    eprint = "1107.5043",
    archivePrefix = "arXiv",
    primaryClass = "astro-ph.SR",
    doi = "10.1088/0004-637X/745/1/70",
    journal = "Astrophys. J.",
    volume = "745",
    pages = "70",
    year = "2012"
}

@ARTICLE{2009CBET.1667....1B,
       author = {{Benetti}, S. and {Valenti}, S. and {Magazzu}, A. and {Harutyunyan}, A.},
        title = "{Supernova 2009H in NGC 1084}",
      journal = {Central Bureau Electronic Telegrams},
         year = 2009,
        month = jan,
       volume = {1667},
        pages = {1},
       adsurl = {https://ui.adsabs.harvard.edu/abs/2009CBET.1667....1B}
}

@article{Inserra:2012gm,
    author = "Inserra, C. and others",
    title = "{The bright Type IIP SN 2009bw, showing signs of interaction}",
    eprint = "1202.0659",
    archivePrefix = "arXiv",
    primaryClass = "astro-ph.SR",
    doi = "10.1111/j.1365-2966.2012.20685.x",
    journal = "Mon. Not. Roy. Astron. Soc.",
    volume = "422",
    pages = "1122--1139",
    year = "2012"
}

@ARTICLE{2009ATel.2106....1I,
       author = {{Immler}, S. and {Russell}, B.~R. and {Brown}, P.~J.},
        title = "{Swift XRT Detection of Supernova 2009dd in X-Rays}",
      journal = {The Astronomer's Telegram},
         year = 2009,
        month = jul,
       volume = {2106},
        pages = {1},
       adsurl = {https://ui.adsabs.harvard.edu/abs/2009ATel.2106....1I}
}

@ARTICLE{2009CBET.1806....1N,
       author = {{Navasardyan}, H. and {Benetti}, S.},
        title = "{Supernova 2009em in NGC 157}",
      journal = {Central Bureau Electronic Telegrams},
         year = 2009,
        month = may,
       volume = {1806},
        pages = {1},
       adsurl = {https://ui.adsabs.harvard.edu/abs/2009CBET.1806....1N}
}

@ARTICLE{2009CBET.1858....1F,
       author = {{Foley}, R.~J.},
        title = "{Supernova 2009gj in NGC 134}",
      journal = {Central Bureau Electronic Telegrams},
         year = 2009,
        month = jun,
       volume = {1858},
        pages = {1},
       adsurl = {https://ui.adsabs.harvard.edu/abs/2009CBET.1858....1F}
}

@article{Elias-Rosa:2011kdh,
    author = "Elias-Rosa, Nancy and others",
    title = "{The Massive Progenitor of the Possible Type II-Linear Supernova 2009hd in Messier 66}",
    eprint = "1108.2645",
    archivePrefix = "arXiv",
    primaryClass = "astro-ph.SR",
    doi = "10.1088/0004-637X/742/1/6",
    journal = "Astrophys. J.",
    volume = "742",
    pages = "6",
    year = "2011"
}

@article{Tak_ts_2015,
   title={SN 2009ib: a Type II-P supernova with an unusually long plateau},
   volume={450},
   ISSN={1365-2966},
   url={http://dx.doi.org/10.1093/mnras/stv857},
   DOI={10.1093/mnras/stv857},
   number={3},
   journal={Monthly Notices of the Royal Astronomical Society},
   publisher={Oxford University Press (OUP)},
   author={Takáts, K. et al},
   year={2015},
   month=may, pages={3137–3154} }

@ARTICLE{2009CBET.1969....2S,
       author = {{Silverman}, J.~M. and {Kandrashoff}, M.~T. and {Filippenko}, A.~V.},
        title = "{Supernova 2009js in NGC 918}",
      journal = {Central Bureau Electronic Telegrams},
         year = 2009,
        month = oct,
       volume = {1969},
        pages = {2},
       adsurl = {https://ui.adsabs.harvard.edu/abs/2009CBET.1969....2S}
}

@article{Fraser:2010jt,
    author = "Fraser, M. and others",
    title = "{SN 2009md: Another faint supernova from a low mass progenitor}",
    eprint = "1011.6558",
    archivePrefix = "arXiv",
    primaryClass = "astro-ph.SR",
    doi = "10.1111/j.1365-2966.2011.19370.x",
    journal = "Mon. Not. Roy. Astron. Soc.",
    volume = "417",
    pages = "1417",
    year = "2011"
}

@ARTICLE{2010ATel.2587....1C,
       author = {{Chomiuk}, Laura and {Soderberg}, Alicia},
        title = "{Radio observations of SN 2010br}",
      journal = {The Astronomer's Telegram},
         year = 2010,
        month = apr,
       volume = {2587},
        pages = {1},
       adsurl = {https://ui.adsabs.harvard.edu/abs/2010ATel.2587....1C}
}

@ARTICLE{2011CBET.2667....1M,
       author = {{Morrell}, N. and {Stritzinger}, M.},
        title = "{Supernova 2011am in NGC 4219 = Psn J12162600-4319200}",
      journal = {Central Bureau Electronic Telegrams},
         year = 2011,
        month = mar,
       volume = {2667},
        pages = {1},
       adsurl = {https://ui.adsabs.harvard.edu/abs/2011CBET.2667....1M}
}

@ARTICLE{Griga2011,
       author = {{Griga}, T. and {Marulla}, A. and {Grenier}, A. and {Sun}, G. and {Gao}, X. and {Lamotte Bailey}, S. and {Koff}, R.~A. and {Mikuz}, H. and {Dintinjana}, B. and {Silverman}, J.~M. and {Cenko}, S.~B. and {Filippenko}, A.~V. and {Li}, W. and {Yamanaka}, M. and {Itoh}, R. and {Arai}, A. and {Nagashima}, M. and {Kajiawa}, K.},
        title = "{Supernova 2011dh in M51 = Psn J13303600+4706330}",
      journal = {Central Bureau Electronic Telegrams},
         year = 2011,
        month = jun,
       volume = {2736},
        pages = {1},
       adsurl = {https://ui.adsabs.harvard.edu/abs/2011CBET.2736....1G}
}

@ARTICLE{2011CBET.2749....1M,
       author = {{Monard}, L.~A.~G. and {Valenti}, S. and {Benetti}, S.},
        title = "{Supernova 2011dq in NGC 337 = Psn J00594775-0734205}",
      journal = {Central Bureau Electronic Telegrams},
         year = 2011,
        month = jun,
       volume = {2749},
        pages = {1},
       adsurl = {https://ui.adsabs.harvard.edu/abs/2011CBET.2749....1M}
}

@article{Eldridge:2013tn,
    author = "Eldridge, John J. and Fraser, Morgan and Smartt, Stephen J. and Maund, Justyn R. and Crockett, R. Mark",
    title = "{The death of massive stars - II. Observational constraints on the progenitors of type Ibc supernovae}",
    eprint = "1301.1975",
    archivePrefix = "arXiv",
    primaryClass = "astro-ph.SR",
    doi = "10.1093/mnras/stt1612",
    journal = "Mon. Not. Roy. Astron. Soc.",
    volume = "436",
    pages = "774",
    year = "2013"
}

@ARTICLE{2011CBET.2946....1M,
       author = {{Monard}, L.~A.~G. and {Milisavljevic}, D. and {Fesen}, R. and {Pickering}, T. and {Romero-Colmenero}, E. and {Turatto}, M. and {Benetti}, S. and {Pastorello}, A. and {Valenti}, S. and {Bufano}, F. and {Tomasella}, L. and {Ryder}, S. and {Soderberg}, A. and {Stockdale}, C. and {van Dyk}, S. and {Immler}, S. and {Weiler}, K. and {Panagia}, N.},
        title = "{Supernova 2011ja in NGC 4945 = Psn J13051112-4931270}",
      journal = {Central Bureau Electronic Telegrams},
         year = 2011,
        month = dec,
       volume = {2946},
        pages = {1},
       adsurl = {https://ui.adsabs.harvard.edu/abs/2011CBET.2946....1M}
}

@ARTICLE{2012CBET.2975....1C,
       author = {{Cao}, Y. and {Kasliwal}, M.~M. and {Wallerstein}, G. and {Ritchey}, A. and {Howell}, D.~A.},
        title = "{Supernova 2012A in NGC 3239}",
      journal = {Central Bureau Electronic Telegrams},
         year = 2012,
        month = jan,
       volume = {2975},
        pages = {1},
       adsurl = {https://ui.adsabs.harvard.edu/abs/2012CBET.2975....1C}
}

@ARTICLE{2012CBET.3054....1F,
       author = {{Fagotti}, P. and {Dimai}, A. and {Quadri}, U. and {Strabla}, L. and {Girelli}, R. and {Quadri}, A. and {Fiorentino}, L. and {Skvarc}, J. and {Masi}, G.},
        title = "{Supernova 2012aw in M95 = PSN J10435372+1140177.}",
      journal = {Central Bureau Electronic Telegrams},
         year = 2012,
        month = mar,
       volume = {3054},
        pages = {1},
       adsurl = {https://ui.adsabs.harvard.edu/abs/2012CBET.3054....1F}
}

@ARTICLE{2012CBET.3105....2M,
       author = {{Marion}, G.~H. and {Milisavljevic}, D. and {Irwin}, J.},
        title = "{Supernova 2012cc in NGC 4419 = PSN J12265681+1502455.}",
      journal = {Central Bureau Electronic Telegrams},
         year = 2012,
        month = may,
       volume = {3105},
        pages = {2},
       adsurl = {https://ui.adsabs.harvard.edu/abs/2012CBET.3105....2M}
}

@article{Barbarino_2015,
   title={SN 2012ec: mass of the progenitor from PESSTO follow-up of the photospheric phase},
   volume={448},
   ISSN={0035-8711},
   url={http://dx.doi.org/10.1093/mnras/stv106},
   DOI={10.1093/mnras/stv106},
   number={3},
   journal={Monthly Notices of the Royal Astronomical Society},
   publisher={Oxford University Press (OUP)},
   author={Barbarino, C. and Dall’Ora, M. and Botticella, M. T. and Valle, M. Della and Zampieri, L. and Maund, J. R. and Pumo, M. L. and Jerkstrand, A. and Benetti, S. and Elias-Rosa, N. and Fraser, M. and Gal-Yam, A. and Hamuy, M. and Inserra, C. and Knapic, C. and LaCluyze, A. P. and Molinaro, M. and Ochner, P. and Pastorello, A. and Pignata, G. and Reichart, D. E. and Ries, C. and Riffeser, A. and Schmidt, B. and Schmidt, M. and Smareglia, R. and Smartt, S. J. and Smith, K. and Sollerman, J. and Sullivan, M. and Tomasella, L. and Turatto, M. and Valenti, S. and Yaron, O. and Young, D.},
   year={2015},
   month=feb, pages={2312–2331} }

@ARTICLE{2012CBET.3263....1N,
       author = {{Nakano}, S. and {Yusa}, T. and {Yoshimoto}, K. and {Tomasella}, L. and {Turatto}, M. and {Pastorello}, A. and {Ochner}, P. and {Cappellaro}, E. and {Takaki}, K. and {Itoh}, R. and {Ueno}, I. and {Urano}, T. and {Moritani}, Y. and {Akitaya}, H. and {Kawabata}, K.~S. and {Yamanaka}, M.},
        title = "{Supernova 2012fh in NGC 3344 = Psn J10433405+2453290}",
      journal = {Central Bureau Electronic Telegrams},
         year = 2012,
        month = oct,
       volume = {3263},
        pages = {1},
       adsurl = {https://ui.adsabs.harvard.edu/abs/2012CBET.3263....1N}
}

@ARTICLE{2013CBET.3422....1B,
       author = {{Blanchard}, P. and {Zheng}, W. and {Cenko}, S.~B. and {Li}, W. and {Filippenko}, A.~V. and {Brimacombe}, J. and {Martignoni}, M. and {Cucchiara}, A. and {Valenti}, S. and {Sand}, D. and {Parrent}, J.~T. and {Graham}, M.~L. and {Howell}, D.~A.},
        title = "{Supernova 2013ab in NGC 5669 = Psn J14324449+0953123}",
      journal = {Central Bureau Electronic Telegrams},
         year = 2013,
        month = feb,
       volume = {3422},
        pages = {1},
       adsurl = {https://ui.adsabs.harvard.edu/abs/2013CBET.3422....1B}
}

@ARTICLE{2013CBET.3440....1N,
       author = {{Nakano}, S. and {Sugano}, M. and {Kadota}, K. and {Benetti}, S. and {Tomasella}, L. and {Pastorello}, A. and {Cappellaro}, E. and {Turatto}, M. and {Ochner}, P.},
        title = "{Supernova 2013am in M65 = Psn J11185695+1303494}",
      journal = {Central Bureau Electronic Telegrams},
         year = 2013,
        month = mar,
       volume = {3440},
        pages = {1},
       adsurl = {https://ui.adsabs.harvard.edu/abs/2013CBET.3440....1N}
}

@ARTICLE{2013CBET.3498....1I,
       author = {{Itagaki}, K. and {Noguchi}, T. and {Nakano}, S. and {Elenin}, L. and {Molotov}, I. and {Moritani}, Y. and {Takaki}, K. and {Kawabata}, K.~S. and {Akitaya}, H. and {Ebisuda}, N. and {Kawaguchi}, K. and {Mori}, K. and {Ohashi}, Y. and {Ueno}, I. and {Sasada}, M. and {Yamanaka}, M. and {Ochner}, P. and {Tomasella}, L. and {Pastorello}, A. and {Benetti}, S. and {Cappellaro}, E. and {Turatto}, M.},
        title = "{Supernova 2013bu in NGC 7331 = Psn J22370217+3424052}",
      journal = {Central Bureau Electronic Telegrams},
         year = 2013,
        month = apr,
       volume = {3498},
        pages = {1},
       adsurl = {https://ui.adsabs.harvard.edu/abs/2013CBET.3498....1I}
}

@ARTICLE{2013CBET.3557....1C,
       author = {{Ciabattari}, F. and {Mazzoni}, E. and {Donati}, S. and {Petroni}, G. and {Foglia}, S. and {Galli}, G. and {Cenko}, S.~B. and {Clubb}, K.~I. and {Zheng}, W. and {Kelly}, P.~L. and {Filippenko}, A.~V. and {Van Dyk}, S.~D.},
        title = "{Supernova 2013df in NGC 4414 = Psn J12262933+3113383}",
      journal = {Central Bureau Electronic Telegrams},
         year = 2013,
        month = jun,
       volume = {3557},
        pages = {1},
       adsurl = {https://ui.adsabs.harvard.edu/abs/2013CBET.3557....1C}
}

@ARTICLE{2013CBET.3565....1C,
       author = {{Carrasco}, F. and {Hamuy}, M. and {Antezana}, R. and {Gonzalez}, L. and {Cartier}, R. and {Forster}, F. and {Silva}, S. and {Ramirez}, R. and {Pignata}, G. and {Apostolovski}, Y. and {Paillas}, E. and {Varela}, S. and {Aros}, F. and {Conuel}, B. and {Folatelli}, G. and {Reichart}, D.~E. and {Haislip}, J.~B. and {Moore}, J.~P. and {LaCluyze}, A.~P. and {Vinko}, J. and {Marion}, G.~H. and {Silverman}, J.~M. and {Wheeler}, J.~C. and {Szalai}, T. and {Quimby}, R.},
        title = "{Supernova 2013dk in NGC 4038 = Psn J12015272-1852183}",
      journal = {Central Bureau Electronic Telegrams},
         year = 2013,
        month = jun,
       volume = {3565},
        pages = {1},
       adsurl = {https://ui.adsabs.harvard.edu/abs/2013CBET.3565....1C}
}

@ARTICLE{2013CBET.3597....1C,
       author = {{Cortini}, G. and {Brimacombe}, J. and {Tomasella}, L. and {Ochner}, P. and {Pastorello}, A. and {Benetti}, S. and {Cappellaro}, E. and {Turatto}, M. and {Farinato}, J. and {Pursimo}, T. and {Dennefeld}, M.},
        title = "{Supernova 2013ee in NGC 3079 = Psn J10015683+5541440}",
      journal = {Central Bureau Electronic Telegrams},
         year = 2013,
        month = jul,
       volume = {3597},
        pages = {1},
       adsurl = {https://ui.adsabs.harvard.edu/abs/2013CBET.3597....1C}
}

@article{Huang:2015kga,
    author = "Huang, Fang and Wang, Xiaofeng and Zhang, Jujia and Brown, Peter J. and Zampieri, Luca and Pumo, Maria Letizia and Zhang, Tianmeng and Chen, Juncheng and Mo, Jun and Zhao, Xulin",
    title = "{SN 2013ej in M74: A Luminous and Fast-declining Type II-P Supernova}",
    eprint = "1504.00446",
    archivePrefix = "arXiv",
    primaryClass = "astro-ph.SR",
    doi = "10.1088/0004-637X/807/1/59",
    journal = "Astrophys. J.",
    volume = "807",
    number = "1",
    pages = "59",
    year = "2015"
}

@ARTICLE{2013CBET.3701....1N,
       author = {{Nakano}, S. and {Kiyota}, S. and {Masi}, Gianluca and {Nocentini}, Francesca and {Schmeer}, Patrick and {Zhang}, J. -J. and {Wang}, X. -F.},
        title = "{Supernova 2013ge in NGC 3287 = Psn J10344846+2139419}",
      journal = {Central Bureau Electronic Telegrams},
         year = 2013,
        month = nov,
       volume = {3701},
        pages = {1},
       adsurl = {https://ui.adsabs.harvard.edu/abs/2013CBET.3701....1N}
}

@ARTICLE{2014CBET.3771....1K,
       author = {{Kim}, H. and {Zheng}, W. and {Li}, W. and {Filippenko}, A.~V. and {Cenko}, S.~B. and {Le Guillou}, L. and {Fleury}, M. and {Baumont}, S. and {Leget}, P. -F. and {Inserra}, C. and {Smartt}, S. and {Smith}, K. and {Young}, D. and {Sullivan}, M. and {Taubenberger}, S. and {Valenti}, S. and {Fraser}, M. and {Yaron}, O. and {Manulis}, I. and {Gal-Yam}, A. and {Knapic}, C. and {Smareglia}, R. and {Molinaro}, M. and {Hsiao}, E.~Y. and {Prieto}, J.~L.},
        title = "{Supernova 2014A in NGC 5054 = Psn J13165936-1637570}",
      journal = {Central Bureau Electronic Telegrams},
         year = 2014,
        month = jan,
       volume = {3771},
        pages = {1},
       adsurl = {https://ui.adsabs.harvard.edu/abs/2014CBET.3771....1K}
}

@ARTICLE{2014CBET.3777....1K,
       author = {{Kim}, M. and {Zheng}, W. and {Li}, W. and {Filippenko}, A.~V. and {Cenko}, S.~B. and {Arbour}, Ron and {Masi}, Gianluca and {Nocentini}, Francesca and {Schmeer}, Patrick and {Zhang}, Jujia and {Want}, Xiaofeng and {Tartaglia}, L. and {Pastorello}, A. and {Benetti}, S. and {Cappellaro}, E. and {Tomasella}, L. and {Ochner}, P. and {Elias-Rosa}, N. and {Turatto}, M.},
        title = "{Supernova 2014C in NGC 7331 = Psn J22370560+3424319}",
      journal = {Central Bureau Electronic Telegrams},
         year = 2014,
        month = jan,
       volume = {3777},
        pages = {1},
       adsurl = {https://ui.adsabs.harvard.edu/abs/2014CBET.3777....1K}
}

@article{Zhang:2018owk,
    author = "Zhang, Jujia and others",
    title = "{Optical Observations of the Young Type Ic Supernova SN 2014L in M99}",
    eprint = "1806.08477",
    archivePrefix = "arXiv",
    primaryClass = "astro-ph.HE",
    doi = "10.3847/1538-4357/aaceaf",
    journal = "Astrophys. J.",
    volume = "863",
    number = "1",
    pages = "109",
    year = "2018"
}

@ARTICLE{2014ATel.6270....1A,
       author = {{Argo}, M. and {Perez-Torres}, M. and {Alberdi}, A. and {Beswick}, R. and {Conway}, J. and {Elias-Rosa}, N. and {Herrero-Illana}, R. and {Kotak}, R. and {Lundqvist}, P. and {Mattila}, S. and {Marcaide}, J. and {Muxlow}, T. and {Marti-Vidal}, I. and {Ramirez-Olivencia}, N. and {Romero-Canizales}, C. and {Stockdale}, C. and {Varenius}, E.},
        title = "{Radio Detection of SN 2014bc in NGC 4258 with eMERLIN}",
      journal = {The Astronomer's Telegram},
         year = 2014,
        month = jun,
       volume = {6270},
        pages = {1},
       adsurl = {https://ui.adsabs.harvard.edu/abs/2014ATel.6270....1A}
}

@ARTICLE{2014CBET.3892....1K,
       author = {{Kumar}, S. and {Zheng}, W. and {Filippenko}, A.~V. and {Masi}, G. and {Zhang}, J. -j. and {Wang}, X. -f.},
        title = "{Supernova 2014bi in NGC 4096 = Psn J12060299+4729335}",
      journal = {Central Bureau Electronic Telegrams},
         year = 2014,
        month = jun,
       volume = {3892},
        pages = {1},
       adsurl = {https://ui.adsabs.harvard.edu/abs/2014CBET.3892....1K}
}

@ARTICLE{2014CBET.3963....1N,
       author = {{Nakano}, S. and {Itagaki}, K. and {Yusa}, T. and {Howerton}, S. and {Elias-Rosa}, N. and {Tartaglia}, L. and {Cappellaro}, E. and {Pastorello}, A. and {Botticella}, M.~T. and {Inserra}, C. and {Maguire}, K. and {Smartt}, S. and {Smith}, K.~W. and {Sullivan}, M. and {Valenti}, S. and {Yaron}, O. and {Young}, D. and {Manulis}, I.},
        title = "{Supernova 2014cx in NGC 337 = Psn J00594783-0734186}",
      journal = {Central Bureau Electronic Telegrams},
         year = 2014,
        month = sep,
       volume = {3963},
        pages = {1},
       adsurl = {https://ui.adsabs.harvard.edu/abs/2014CBET.3963....1N}
}

@ARTICLE{2014CBET.3977....1M,
       author = {{Monard}, L.~A.~G. and {Kneip}, R. and {Brimacombe}, J. and {Sato}, H. and {Childress}, M. and {Zhou}, G. and {Scalzo}, R. and {Yuan}, F. and {Zhang}, B. and {Ruiter}, A. and {Seitenzahl}, I. and {Schmidt}, B. and {Tucker}, B.},
        title = "{Supernova 2014df in NGC 1448 = Psn J03442399-4440081}",
      journal = {Central Bureau Electronic Telegrams},
         year = 2014,
        month = sep,
       volume = {3977},
        pages = {1},
       adsurl = {https://ui.adsabs.harvard.edu/abs/2014CBET.3977....1M}
}

@ARTICLE{2015CBET.4087....1Y,
       author = {{Yusa}, T. and {Buczynski}, D. and {Noguchi}, T. and {Nakano}, S. and {Kiyota}, S. and {Masi}, G. and {Ayani}, K. and {Foley}, R.~J. and {Zheng}, W. and {Filippenko}, A.~V. and {Van Dyk}, S.~D.},
        title = "{Supernova 2015G in NGC 6951 = Psn J20372558+6607115}",
      journal = {Central Bureau Electronic Telegrams},
         year = 2015,
        month = apr,
       volume = {4087},
        pages = {1},
       adsurl = {https://ui.adsabs.harvard.edu/abs/2015CBET.4087....1Y}
}

@ARTICLE{2016ATel.8566....1B,
       author = {{Bock}, G. and {Shappee}, B.~J. and {Stanek}, K.~Z. and {Prieto}, J.~L. and {Kochanek}, C.~S. and {Holoien}, T.~W. -S. and {Brown}, J.~S. and {Godoy-Rivera}, D. and {Basu}, U. and {Bersier}, D. and {Dong}, Subo and {Chen}, Ping and {Brimacombe}, J. and {Masi}, G. and {Kiyota}, S.},
        title = "{ASASSN-16at: Discovery of A Probable Nearby Supernova in UGC 08041}",
      journal = {The Astronomer's Telegram},
         year = 2016,
        month = jan,
       volume = {8566},
        pages = {1},
       adsurl = {https://ui.adsabs.harvard.edu/abs/2016ATel.8566....1B}
}

@ARTICLE{2016ATel.8662....1B,
       author = {{Brown}, Peter J. and {Landez}, Nancy J. and {Beeny}, Britton A.},
        title = "{Swift/UVOT Observations of SN2016adj in NGC5128}",
      journal = {The Astronomer's Telegram},
         year = 2016,
        month = feb,
       volume = {8662},
        pages = {1},
       adsurl = {https://ui.adsabs.harvard.edu/abs/2016ATel.8662....1B}
}

@ARTICLE{2016TNSCR.296....1J,
       author = {{Johansson}, J. and {Magee}, M. and {Bar}, I. and {Yaron}, O.},
        title = "{PESSTO Transient Classification Report for 2016-04-16}",
      journal = {Transient Name Server Classification Report},
         year = 2016,
        month = apr,
       volume = {2016-296},
        pages = {1},
       adsurl = {https://ui.adsabs.harvard.edu/abs/2016TNSCR.296....1J}
}

@ARTICLE{2016ATel.9091....1B,
       author = {{Bock}, G. and {Dong}, Subo and {Kochanek}, C.~S. and {Stanek}, K.~Z. and {Brown}, J.~S. and {Holoien}, T.~W. -S. and {Godoy-Rivera}, D. and {Basu}, U. and {Shappee}, B.~J. and {Prieto}, J.~L. and {Bersier}, D. and {Chen}, Ping and {Brimacombe}, J.},
        title = "{ASASSN-16fq: Discovery of A Probable Supernova in M66}",
      journal = {The Astronomer's Telegram},
         year = 2016,
        month = may,
       volume = {9091},
        pages = {1},
       adsurl = {https://ui.adsabs.harvard.edu/abs/2016ATel.9091....1B}
}

@ARTICLE{2023MNRAS.523.6048S,
       author = {{Shahbandeh}, Melissa and {Sarangi}, Arkaprabha and {Temim}, Tea and {Szalai}, Tam{\'a}s and {Fox}, Ori D. and {Tinyanont}, Samaporn and {Dwek}, Eli and {Dessart}, Luc and {Filippenko}, Alexei V. and {Brink}, Thomas G. and {Foley}, Ryan J. and {Jencson}, Jacob and {Pierel}, Justin and {Zs{\'\i}ros}, Szanna and {Rest}, Armin and {Zheng}, WeiKang and {Andrews}, Jennifer and {Clayton}, Geoffrey C. and {De}, Kishalay and {Engesser}, Michael and {Gezari}, Suvi and {Gomez}, Sebastian and {Gonzaga}, Shireen and {Johansson}, Joel and {Kasliwal}, Mansi and {Lau}, Ryan and {De Looze}, Ilse and {Marston}, Anthony and {Milisavljevic}, Dan and {O'Steen}, Richard and {Siebert}, Matthew and {Skrutskie}, Michael and {Smith}, Nathan and {Strolger}, Lou and {Van Dyk}, Schuyler D. and {Wang}, Qinan and {Williams}, Brian and {Williams}, Robert and {Xiao}, Lin and {Yang}, Yi},
        title = "{JWST observations of dust reservoirs in type IIP supernovae 2004et and 2017eaw}",
      journal = {\mnras},
         year = 2023,
        month = aug,
       volume = {523},
       number = {4},
        pages = {6048-6060},
          doi = {10.1093/mnras/stad1681},
archivePrefix = {arXiv},
       eprint = {2301.10778},
 primaryClass = {astro-ph.HE},
       adsurl = {https://ui.adsabs.harvard.edu/abs/2023MNRAS.523.6048S}
}

@ARTICLE{2017TNSCR.881....1L,
       author = {{Lyman}, J. and {Homan}, D. and {Magee}, M. and {Yaron}, O.},
        title = "{ePESSTO Transient Classification Report for 2017-08-17}",
      journal = {Transient Name Server Classification Report},
         year = 2017,
        month = aug,
       volume = {2017-881},
        pages = {1},
       adsurl = {https://ui.adsabs.harvard.edu/abs/2017TNSCR.881....1L}
}

@ARTICLE{2017TNSCR.964....1O,
       author = {{Onori}, F.},
        title = "{NUTS Transient Classification Report for 2017-09-03}",
      journal = {Transient Name Server Classification Report},
         year = 2017,
        month = sep,
       volume = {2017-964},
        pages = {1},
       adsurl = {https://ui.adsabs.harvard.edu/abs/2017TNSCR.964....1O}
}

@article{Dastidar:2025yxz,
    author = "Dastidar, R. and others",
    title = "{SN 2018is: A low-luminosity Type IIP supernova with narrow hydrogen emission lines at early phases}",
    eprint = "2501.01530",
    archivePrefix = "arXiv",
    primaryClass = "astro-ph.HE",
    doi = "10.1051/0004-6361/202452507",
    journal = "Astron. Astrophys.",
    volume = "694",
    pages = "A260",
    year = "2025"
}

@article{Jacobson-Galan:2020gyk,
    author = "Jacobson-Gal{\'a}n, Wynn V. and others",
    title = "{SN 2019ehk: A Double-Peaked Ca-rich Transient with Luminous X-ray Emission and Shock-Ionized Spectral Features}",
    eprint = "2005.01782",
    archivePrefix = "arXiv",
    primaryClass = "astro-ph.HE",
    doi = "10.3847/1538-4357/ab9e66",
    journal = "Astrophys. J.",
    volume = "898",
    number = "2",
    pages = "166",
    year = "2020"
}

@ARTICLE{2020TNSCR..90....1S,
       author = {{Siebert}, M.~R. and {Kilpatrick}, C.~D. and {Foley}, R.~J. and {Cartier}, R.},
        title = "{UCSC Transient Classification Report for 2020-01-09}",
      journal = {Transient Name Server Classification Report},
         year = 2020,
        month = jan,
       volume = {2020-90},
        pages = {1},
       adsurl = {https://ui.adsabs.harvard.edu/abs/2020TNSCR..90....1S}
}

@ARTICLE{2020TNSCR.657....1K,
       author = {{Kawabata}, M.},
        title = "{Transient Classification Report for 2020-02-27}",
      journal = {Transient Name Server Classification Report},
         year = 2020,
        month = feb,
       volume = {2020-657},
        pages = {1},
       adsurl = {https://ui.adsabs.harvard.edu/abs/2020TNSCR.657....1K}
}

@ARTICLE{2020ATel14226....1K,
       author = {{Kundu}, E. and {Ryder}, S.~D. and {Filipovic}, M.~D. and {Anderson}, G. and {Stockdale}, C. and {Maeda}, K. and {Renaud}, M. and {Kotak}, R.},
        title = "{Radio observations of SN 2020zbv}",
      journal = {The Astronomer's Telegram},
         year = 2020,
        month = nov,
       volume = {14226},
        pages = {1},
       adsurl = {https://ui.adsabs.harvard.edu/abs/2020ATel14226....1K}
}

@ARTICLE{2020TNSCR3777....1L,
       author = {{Lundquist}, M. and {Andrews}, J. and {Valenti}, S. and {Sand}, D.~J. and {Wyatt}, S. and {Amaro}, R. and {Jencson}, J. and {Dong}, Y. and {Davis}, S. and {Janzen}, D.},
        title = "{DLT40 Transient Classification Report for 2020-12-13}",
      journal = {Transient Name Server Classification Report},
         year = 2020,
        month = dec,
       volume = {2020-3777},
        pages = {1},
       adsurl = {https://ui.adsabs.harvard.edu/abs/2020TNSCR3777....1L}
}

@ARTICLE{2022TNSCR.233....1B,
       author = {{Bostroem}, K.~A. and {Andrews}, J. and {Sand}, D.~J. and {Wyatt}, S. and {Lundquist}, M. and {Jencson}, J. and {Dong}, Y. and {Janzen}, D. and {Hosseinzadeh}, G. and {Pearson}, J. and {Valenti}, S. and {Retamal}, N.},
        title = "{DLT40 Transient Classification Report for 2022-01-28}",
      journal = {Transient Name Server Classification Report},
         year = 2022,
        month = jan,
       volume = {2022-233},
        pages = {1},
       adsurl = {https://ui.adsabs.harvard.edu/abs/2022TNSCR.233....1B}
}

@ARTICLE{2022TNSCR1261....1G,
       author = {{Grzegorzek}, J.},
        title = "{PSH Transient Classification Report for 2022-05-11}",
      journal = {Transient Name Server Classification Report},
         year = 2022,
        month = may,
       volume = {2022-1261},
        pages = {1},
       adsurl = {https://ui.adsabs.harvard.edu/abs/2022TNSCR1261....1G}
}

@ARTICLE{2022TNSCR1820....1B,
       author = {{Bayer}, J. and {Wang}, H. and {Huber}, S. and {Vogl}, C. and {Taubenberger}, S. and {Kressierer}, S. and {Cudmani}, M.~G. and {Holas}, A. and {Benetti}, S. and {Cappellaro}, E. and {Pastorello}, A. and {Reguitti}, A. and {Tomasella}, L. and {Pignata}, G.},
        title = "{Padova-Asiago Transient Classification Report for 2022-06-29}",
      journal = {Transient Name Server Classification Report},
         year = 2022,
        month = jun,
       volume = {2022-1820},
        pages = {1},
       adsurl = {https://ui.adsabs.harvard.edu/abs/2022TNSCR1820....1B}
}

@ARTICLE{2022TNSCR2308....1H,
       author = {{Hinds}, K. and {Perley}, D.},
        title = "{ZTF Transient Classification Report for 2022-08-12}",
      journal = {Transient Name Server Classification Report},
         year = 2022,
        month = aug,
       volume = {2022-2308},
        pages = {1},
       adsurl = {https://ui.adsabs.harvard.edu/abs/2022TNSCR2308....1H}
}

@ARTICLE{2022TNSCR3048....1B,
       author = {{Balcon}, C.},
        title = "{Transient Classification Report for 2022-10-19}",
      journal = {Transient Name Server Classification Report},
         year = 2022,
        month = oct,
       volume = {2022-3048},
        pages = {1},
       adsurl = {https://ui.adsabs.harvard.edu/abs/2022TNSCR3048....1B}
}

@ARTICLE{2023TNSCR.990....1K,
       author = {{Karambelkar}, V.},
        title = "{ZTF Transient Classification Report for 2023-05-01}",
      journal = {Transient Name Server Classification Report},
         year = 2023,
        month = may,
       volume = {2023-990},
        pages = {1},
       adsurl = {https://ui.adsabs.harvard.edu/abs/2023TNSCR.990....1K}
}

@ARTICLE{2023TNSCR1164....1P,
       author = {{Perley}, D. and {Gal-Yam}, A.},
        title = "{Transient Classification Report for 2023-05-19}",
      journal = {Transient Name Server Classification Report},
         year = 2023,
        month = may,
       volume = {2023-1164},
        pages = {1},
       adsurl = {https://ui.adsabs.harvard.edu/abs/2023TNSCR1164....1P}
}

@ARTICLE{2023TNSCR1699....1T,
       author = {{Tucker}, M.~A.},
        title = "{SCAT Transient Classification Report for 2023-07-17}",
      journal = {Transient Name Server Classification Report},
         year = 2023,
        month = jul,
       volume = {2023-1699},
        pages = {1},
       adsurl = {https://ui.adsabs.harvard.edu/abs/2023TNSCR1699....1T}
}

@ARTICLE{2023TNSCR2224....1B,
       author = {{Balcon}, C.},
        title = "{Transient Classification Report for 2023-09-10}",
      journal = {Transient Name Server Classification Report},
         year = 2023,
        month = sep,
       volume = {2023-2224},
        pages = {1},
       adsurl = {https://ui.adsabs.harvard.edu/abs/2023TNSCR2224....1B}
}

@ARTICLE{2023TNSCR2786....1B,
       author = {{Balcon}, C.},
        title = "{Transient Classification Report for 2023-10-31}",
      journal = {Transient Name Server Classification Report},
         year = 2023,
        month = oct,
       volume = {2023-2786},
        pages = {1},
       adsurl = {https://ui.adsabs.harvard.edu/abs/2023TNSCR2786....1B}
}

@ARTICLE{2024TNSCR.284....1B,
       author = {{Balcon}, C.},
        title = "{Transient Classification Report for 2024-01-29}",
      journal = {Transient Name Server Classification Report},
         year = 2024,
        month = jan,
       volume = {2024-284},
        pages = {1},
       adsurl = {https://ui.adsabs.harvard.edu/abs/2024TNSCR.284....1B}
}

@ARTICLE{2024TNSCR2403....1A,
       author = {{Andrews}, J.~E. and {Bostroem}, K.~A. and {Valenti}, S. and {Sand}, D.~J. and {Jha}, S. and {Lundquist}, M. and {Hoang}, E. and {Shrestha}, M. and {Dong}, Y. and {Janzen}, D. and {Hosseinzadeh}, G. and {Pearson}, J. and {Meza}, N. and {Metha}, D. and {Ravi}, A. and {Martas}, A.},
        title = "{DLT40 Transient Classification Report for 2024-07-12}",
      journal = {Transient Name Server Classification Report},
         year = 2024,
        month = jul,
       volume = {2024-2403},
        pages = {1},
       adsurl = {https://ui.adsabs.harvard.edu/abs/2024TNSCR2403....1A}
}

@ARTICLE{2024TNSCR2571....1B,
       author = {{Bostroem}, K.~A. and {Kyer}, R. and {Strader}, J. and {Shrestha}, M. and {Newsome}, M. and {Terreran}, G. and {McCully}, C. and {Sand}, D.~J. and {Jha}, S. and {Andrews}, J.~E. and {Valenti}, S. and {Lundquist}, M. and {Hoang}, E. and {Dong}, Y. and {Janzen}, D. and {Hosseinzadeh}, G. and {Pearson}, J. and {Meza}, N. and {Metha}, D. and {Ravi}, A. and {Martas}, A. and {Howell}, D.~A. and {Andrews}, M. and {Farah}, J. and {Gonzalez}, E.~P.},
        title = "{DLT40 Transient Classification Report for 2024-07-24}",
      journal = {Transient Name Server Classification Report},
         year = 2024,
        month = jul,
       volume = {2024-2571},
        pages = {1},
       adsurl = {https://ui.adsabs.harvard.edu/abs/2024TNSCR2571....1B}
}

@ARTICLE{2025TNSCR2281....1H,
       author = {{Hoogendam}, W.},
        title = "{SCAT Transient Classification Report for 2025-06-18}",
      journal = {Transient Name Server Classification Report},
         year = 2025,
        month = jun,
       volume = {2025-2281},
        pages = {1},
       adsurl = {https://ui.adsabs.harvard.edu/abs/2025TNSCR2281....1H}
}

@ARTICLE{2025TNSCR2547....1S,
       author = {{Strader}, J.},
        title = "{Transient Classification Report for 2025-07-03}",
      journal = {Transient Name Server Classification Report},
         year = 2025,
        month = jul,
       volume = {2025-2547},
        pages = {1},
       adsurl = {https://ui.adsabs.harvard.edu/abs/2025TNSCR2547....1S}
}

@misc{suppmat,
  key  = {Supplemental Material},
  note = {See Supplemental Material at [URL] for additional details.}
}

\onecolumngrid
\appendix

\setcounter{equation}{0}
\setcounter{figure}{0}
\setcounter{table}{0}
\setcounter{page}{1}
\makeatletter
\renewcommand{\theequation}{S\arabic{equation}}
\renewcommand{\thefigure}{S\arabic{figure}}
\renewcommand{\thepage}{S\arabic{page}}
\renewcommand{\thetable}{S\arabic{table}}

\begin{center}
\textbf{\large Supplemental Material\\[0.5ex]
{\em Detecting light axions from supernovae in nearby galaxies}}
\end{center}

\bigskip

In the Supplemental Material (SM), we present the supernova catalog used in our analysis, provide fitting formulas for axion production in the supernova core, describe the models adopted for magnetic fields in the various astrophysical environments considered, {and discuss details of the computation of the sensitivities reported in the main text, as well as potential sources of uncertainty affecting them.} We also compare our results with previous literature.

\section{Supernova catalog}
\label{App:SNcatalog}
{In Table~\ref{tab:A1} we present the list of supernovae (SNe) employed in our analysis taken from the Transient Name Server catalog~\cite{2021AAS...23742305G}.
We include only SNe that are unambiguously identified as core-collapse events and that occurred during the \emph{Fermi}-LAT operational period~(2008–2025). In the spirit of this work, the goal here is to estimate how many SN events in the nearby Universe can be detected with current facilities over the typical lifetime of a gamma-ray mission. In the Table, we also quote the distances of the galaxies hosting the SN event. 
Following Ref.~\cite{Nakamura:2016thx}, we use the catalog in Ref.~\cite{Kennicutt:2008ce} for distances $\lesssim 10\,\rm{Mpc}$. For host galaxies at distances $d \gtrsim 10\,\rm{Mpc}$, we refer to Ref.~\cite{Tully:2013wqa} by employing the interface provided by Ref.~\cite{Ochsenbein:2000th}.}

\renewcommand{\arraystretch}{1.5}
\setlength{\LTcapwidth}{\textwidth}
\begin{longtable}{c c c c c}

  \caption{List of observed core-collapse SNe between 2008 and 2025 taken from the Transient Name Server catalog~\cite{2021AAS...23742305G}. The Table reports the supernova name, SN type, distance (in Mpc), name of the host galaxy, and the references confirming each event. We use the SN distances reported in Ref.~\cite{Kennicutt:2008ce} for host galaxies at $d\lesssim 10\,\rm{Mpc}$ and Ref.~\cite{Ochsenbein:2000th} for $d \gtrsim 10\,\rm{Mpc}$.
  \\
  }\\
  \hline
  Name & Type & Distance & Host & References \\
   & & (Mpc) & Galaxy & \\
  \hline
  \hline
  \endfirsthead

  \hline
  Name & Type & Distance & Host & References \\
   & & (Mpc) & Galaxy& \\
  \hline
  \hline
  \endhead

  \hline
  \endfoot
     SN 2008S & SN IIn &  5.9 & NGC6946 & \cite{2008CBET.1381....1W} \\
     SN 2008ax & SN IIP & 8.0 & NGC4490 & \cite{2008CBET.1280....1M} \\
     SN 2008bk & SN IIP & 3.91 & NGC 7793 & \cite{2008ATel.1452....1S} \\
    SN 2008iz &	SN II &	3.53&	NGC 3034 & \cite{Brunthaler:2010bm}	 \\
    SN 2008jb &	SN II &	9.6 &	ESO 302-14	&  \cite{Prieto:2011tb} \\
    SN 2009H &	SN II &	20.89  &	NGC 1084 & \cite{2009CBET.1667....1B}	\\
    SN 2009bw &	SN II &	9.46&	UGC 2890& \cite{Inserra:2012gm} \\
     SN 2009dd & SN II & 14.45 & NGC4088 & \cite{2009ATel.2106....1I} \\
     SN 2009em&	SN Ic&	12.08&	NGC 157& \cite{2009CBET.1806....1N}\\
     SN 2009gj &	SN IIb &	19.95 &	NGC 134 & \cite{2009CBET.1858....1F} \\
     SN 2009hd &	SN II	&	10.05 &	NGC 3627 & \cite{Elias-Rosa:2011kdh}\\
     SN 2009ib &	SN IIP  &	12.59 &	NGC 1559 & \cite{Tak_ts_2015} \\
     SN 2009js&	SN II&	15.92 &	NGC 918& \cite{2009CBET.1969....2S}\\
     SN 2009md &	SN II &	20.8&	NGC 3389& \cite{Fraser:2010jt}\\
     SN 2010br &	SN Ib/c &	11.02  &	NGC 4051 & \cite{2010ATel.2587....1C}	\\
     SN 2011am &	SN Ib &	20.2 &	NGC 4219& \cite{2011CBET.2667....1M}	\\
     SN 2011dh & SN IIP &  8.0 & NGC 5194  & \cite{Griga2011}
     \\
     SN 2011dq &	SN II&18.88&	NGC 337& \cite{2011CBET.2749....1M}\\
     SN 2011hp &	SN Ic&	20.2&	NGC 4219&\cite{Eldridge:2013tn} \\
     2011ja & SN IIP &   3.6& NGC 4945  & \cite{2011CBET.2946....1M}
     \\
     SN 2012A &	SN II &	8.3&	NGC 3239 & \cite{2012CBET.2975....1C}
     \\
     SN 2012aw &	SN IIP &	10.0	& NGC 3351 & \cite{2012CBET.3054....1F}
     \\
     SN 2012cc&	SN II&	13.49&	NGC 4419& \cite{2012CBET.3105....2M}\\
     SN 2012ec &	SN IIP	& 20.89&	NGC 1084 & \cite{Barbarino_2015}
     \\
     SN 2012fh &	SN Ib/c	&	6.6&	NGC3344 & \cite{2012CBET.3263....1N}
     \\
     SN 2013ab&	SN II&	15.630&	NGC 5669& \cite{2013CBET.3422....1B} \\
     SN 2013am &	SN IIP	&	8.9&	M65 & \cite{2013CBET.3440....1N}
     \\
     SN 2013bu&	SN II&	13.87&	NGC 7331& \cite{2013CBET.3498....1I}\\
     SN 2013df&	SN II&	17.86&	NGC 4414& \cite{2013CBET.3557....1C}\\
     SN 2013dk	&SN Ic&	20.610&	NGC 4038& \cite{2013CBET.3565....1C} \\
     SN 2013ee&	SN II&	20.610&	NGC 3079& \cite{2013CBET.3597....1C}\\
     SN 2013ej &	SN II & 7.3 &	NGC 628 & \cite{Huang:2015kga}
     \\
     SN 2013ge &	SN Ib &	14.79 	& NGC 3287 & \cite{2013CBET.3701....1N} \\
     SN 2014A &	SN II &18.2&	NGC 5054 &\cite{2014CBET.3771....1K} \\
     SN 2014C &	SN Ib &13.87 &	NGC 7331 & \cite{2014CBET.3777....1K} \\
     SN 2014L&	SN Ic&	13.870&	NGC 4254&\cite{Zhang:2018owk} \\
     SN 2014bc &	SN IIP &7.98&	NGC 4258 & \cite{2014ATel.6270....1A} \\
     SN 2014bi &	SN II &	8.3&	NGC 4069 & \cite{2014CBET.3892....1K} \\
     SN 2014cx &	SN II &	18.88 & NGC 337 &  \cite{2014CBET.3963....1N} \\
     SN 2014df &	SN Ib &	17.22 &	NGC 1448 & \cite{2014CBET.3977....1M} \\
     SN 2015G&	SN Ibn&	18.79&	NGC 6951&\cite{2015CBET.4087....1Y} \\
     SN 2016X &	SN IIP	&	14.45 &	UGC 08041 & \cite{2016ATel.8566....1B} \\
     SN 2016adj &	SN IIb &	3.6&	NGC 5128 & \cite{2016ATel.8662....1B} \\
     SN 2016bmi &	SN IIP&	19.77 &	IC4721& \cite{2016TNSCR.296....1J}	\\
     SN 2016cok &	SN IIP &10.05	&NGC 3627  & \cite{2016ATel.9091....1B} \\
     SN 2016ija &	SN II &	20.04 &	NGC1532 & \cite{Tartaglia:2017yym} \\
     SN 2017eaw	& SN IIP	& 5.9&	NGC 6946 & \cite{2023MNRAS.523.6048S} \\
     SN 2017gaw&	SN II&	16.9&	UGC807& \cite{2017TNSCR.881....1L}\\
     SN 2017gkk	& SN IIb &	19.77 &	NGC2748 & \cite{2017TNSCR.964....1O} \\
     SN 2018is&	SN II&	18.2&	NGC 5054& \cite{Dastidar:2025yxz} \\
     SN 2019ehk&	SN Ib&	13.93&	NGC 4321&\cite{Jacobson-Galan:2020gyk} \\
     SN 2020oi&	SN Ic	& 13.93&	NGC 4321 & \cite{2020TNSCR..90....1S} \\
     SN 2020dpw&	SN IIP&	18.79&	NGC 6952&\cite{2020TNSCR.657....1K} \\
     SN 2020zbv &	SN IIP &	17.22	& NGC1448 & \cite{2020ATel14226....1K} \\
     SN 2020acli&	SN IIn-pec&	2.0 &	NGC 300& \cite{2020TNSCR3777....1L}\\
     SN 2022ame &	SN II &	13.61	&NGC 1255 & \cite{2022TNSCR.233....1B} \\
     SN 2022jli&	SN Ic&	12.08 &	NGC 157& \cite{2022TNSCR1261....1G} \\
     SN 2022myz&	SN I &	20.14 &	NGC 4217&\cite{2022TNSCR1820....1B} \\
    SN 2022oey&	SN II &	14.39 &	UGC 02855& \cite{2022TNSCR2308....1H} \\
    SN 2022xxf &	SN Ic-BL & 19.14 &	NGC 3705 & \cite{2022TNSCR3048....1B} \\
     SN 2023hlf &	SN II &17.86 &	NGC 4414 & \cite{2023TNSCR.990....1K} \\
     SN 2023ixf&	SN II	&16.95 &	NGC 545& \cite{2023TNSCR1164....1P}\\
     SN 2023mut &	SN IIb &	8.1 &	UGC 03174 & \cite{2023TNSCR1699....1T} \\
     SN 2023rve &	SN II &16.0  &	NGC 1097 & \cite{2023TNSCR2224....1B} \\
     SN 2023wcr&	SN II&	14.06 &	NGC4346& \cite{2023TNSCR2786....1B} \\
     SN 2024bch&	SN II&	18.11 &	NGC 3206& \cite{2024TNSCR.284....1B} \\
     SN 2024phv &	SN II &	19.23	& NGC 3969 & \cite{2024TNSCR2403....1A} \\
     SN 2024pxg & SN II&	12.36 &  NGC6221 &\cite{2024TNSCR2571....1B}\\
     SN 2025nqb&	SN Ib&	14.72&	NGC 4496B& \cite{2025TNSCR2281....1H} \\
     SN 2025pht	&SN IIP&	17.46&	NGC 1637&   ~\cite{2025TNSCR2547....1S} 
     \label{tab:A1}
\end{longtable}

\section{SN ALP spectrum}
\label{App:SNspectrum}

In this work we calculate the SN ALP fluxes using as benchmark the 1D spherical symmetric {\tt GARCHING} group's SN model SFHo-s18.8 provided in Ref.~\cite{SNarchive} and based on the neutrino-hydrodynamics code 
{\tt PROMETHEUS-VERTEX}~\cite{Rampp:2002bq}. The simulation employs the SFHo Equation of State (EoS)~\cite{Hempel:2009mc,Steiner:2012rk} and is launched from a stellar progenitor with mass $18.8~M_\odot$~\cite{Sukhbold:2017cnt}, leading to neutron star (NS) with baryonic mass $1.35~M_\odot$. This model corresponds to the cold model used in Refs.~\cite{Manzari:2024jns,Fiorillo:2025gnd}, leading to the most conservative results.
We consider SN ALP production via the Lagrangian
\begin{equation}
{\mathcal L}_{aN}=
\frac{\partial_\mu a}{2m_N}\sum_{N=p,n}
g_{aN}{\bar N}\gamma^\mu\gamma_5 N \,\ ,
\label{eq:nucllagr}
\end{equation}
from both nucleon-nucleon bremsstrahlung ($NN$) and Compton pionic processes ($\pi N$).
In order to characterize the resulting SN ALP spectra, we integrated in time the spectra used in Ref.~\cite{Lella:2024hfk} from the SFHo-s18.8 model, in the time window [1;8] seconds. For the time-integrated spectra we provide as analytical fit for the bremsstrahlung production

\begin{equation}
    \frac{dN}{dE} = C 
    \left(\frac{g_{a p}}{10^{-9}}\right)^{2}
    \left(\frac{E}{E_0}\right)^{\beta}
    \exp\left[-\frac{(\beta + 1)E}{E_0}\right],
    \label{eq:bremsfit}
\end{equation}
with {$C = 1.22 \times 10^{55}~\text{MeV}^{-1}$, $E_0 = 67.7~\text{MeV}$, and $\beta = 1.3$}.\,The $NN$ spectrum computed here includes all the corrections discussed in Refs.~\cite{Carenza:2019pxu}. Similar results would be obtained using the parametric treatment proposed by Ref.~\cite{Fiorillo:2025gnd}, where all nuclear uncertainties are encoded in the nucleon spin-flip rate, which is quite consistent with Ref.~\cite{Carenza:2019pxu}. 

\begin{figure}
    \centering
    \includegraphics[width=0.7\linewidth]{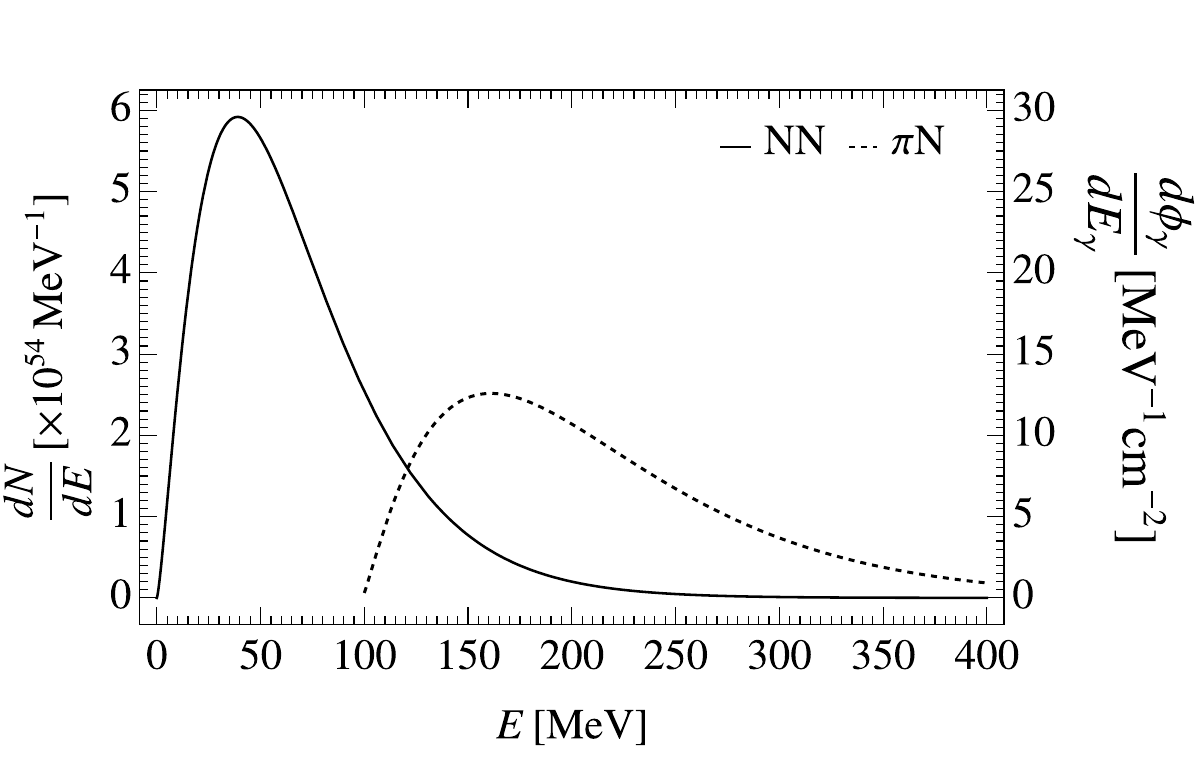}
    \caption{Time-integrated ALP emission spectrum from $NN$ bremsstrahlung (solid line), and pionic processes (dashed line). These spectra are obtained from fitting formulas in Eqs.~(\ref{eq:bremsfit}) and~(\ref{eq:piofit}) by setting $g_{ap} = 10^{-9}$ and $g_{an} = 0$.
    On the right axis the corresponding values of the photon flux from M82 reaching the Earth by considering ALP--photon conversion in the subhalo and along the line of sight that yield the average conversion probability. The ALP--photon conversion probability is estimated at $g_{a\gamma} = 10^{-12}\,\rm{GeV}^{-1}$.}
    \label{fig:plotA1}
\end{figure}

Conversely, if pions are present in the SN core, the resulting time-integrated spectrum  for
the Compton pionic processes  is given by
\begin{equation}
    \frac{dN}{dE} = C_{\pi} 
    \left(\frac{g_{a p}}{10^{-9}}\right)^{2}
    \left(\frac{E - \omega_c}{E_{0\pi}}\right)^{\beta_{\pi}}
    \exp\left[-\frac{(\beta_{\pi} + 1)(E - \omega_c)}{E_{0\pi}}\right],
    \label{eq:piofit}
\end{equation}
with {$C_{\pi} = 4.08 \times 10^{54}~\text{MeV}^{-1}$, $E_{0\pi} = 115.8~\text{MeV}$, $\beta_{\pi} = 1.15$, and $\omega_c = 99.1~\text{MeV}$.}
The total spectrum with these two components is shown in  Fig.~\ref{fig:plotA1}.

\section{Magnetic fields in host galaxies and conversion probabilities}
\label{App:MagneticFields}

In this analysis, we employ publicly available simulation data from the \textsc{IllustrisTNG} project to model the magnetic field and the electron number density of the three hots under study, {namely M82, NGC 6946, Virgo Cluster}~\cite{Nelson:2018uso}. 
The \textsc{IllustrisTNG} project consists of a series of cosmological magnetohydrodynamical simulations of galaxy formation, which also aims to provide predictions for both current and future observational programs. 
For the environments investigated in this work, we used data from different simulation volumes, selecting in each case the highest {space} resolution realizations available. 
The subhalos included in our analysis were identified by searching for candidates capable of reproducing the observed total stellar mass and star-formation rate (SFR) of the three hosts, following Ref.~\cite{Ning:2024eky}. 
The relevant properties of the selected subhalos for the three environments are reported in Table~\ref{tab:A2}.
We highlight that the selected subhalos present physical extensions consistent with those of the corresponding host environments. 

\begin{figure}[t!]
    \centering
    \includegraphics[width=0.495\textwidth]{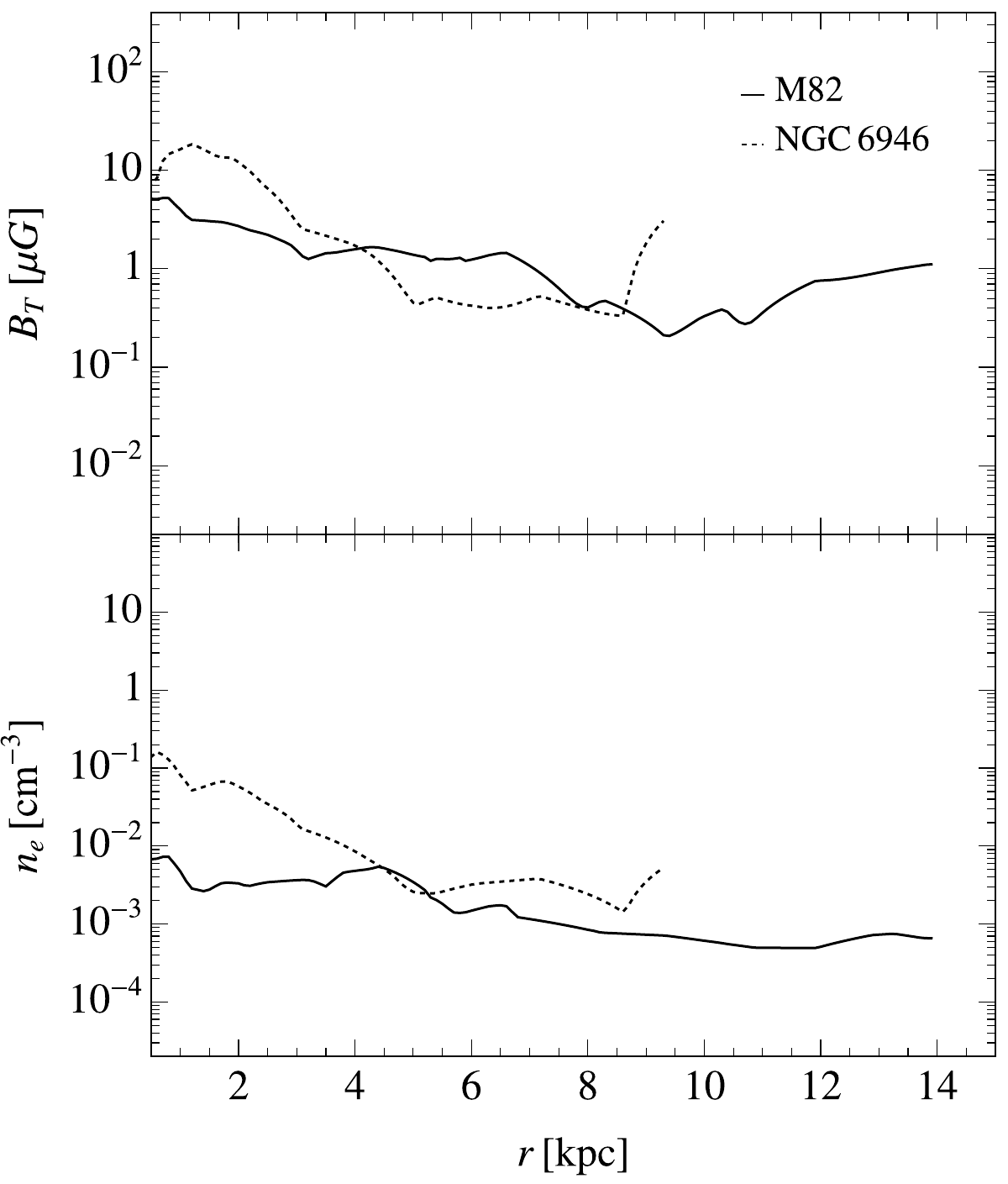}
    \includegraphics[width=0.495\textwidth]{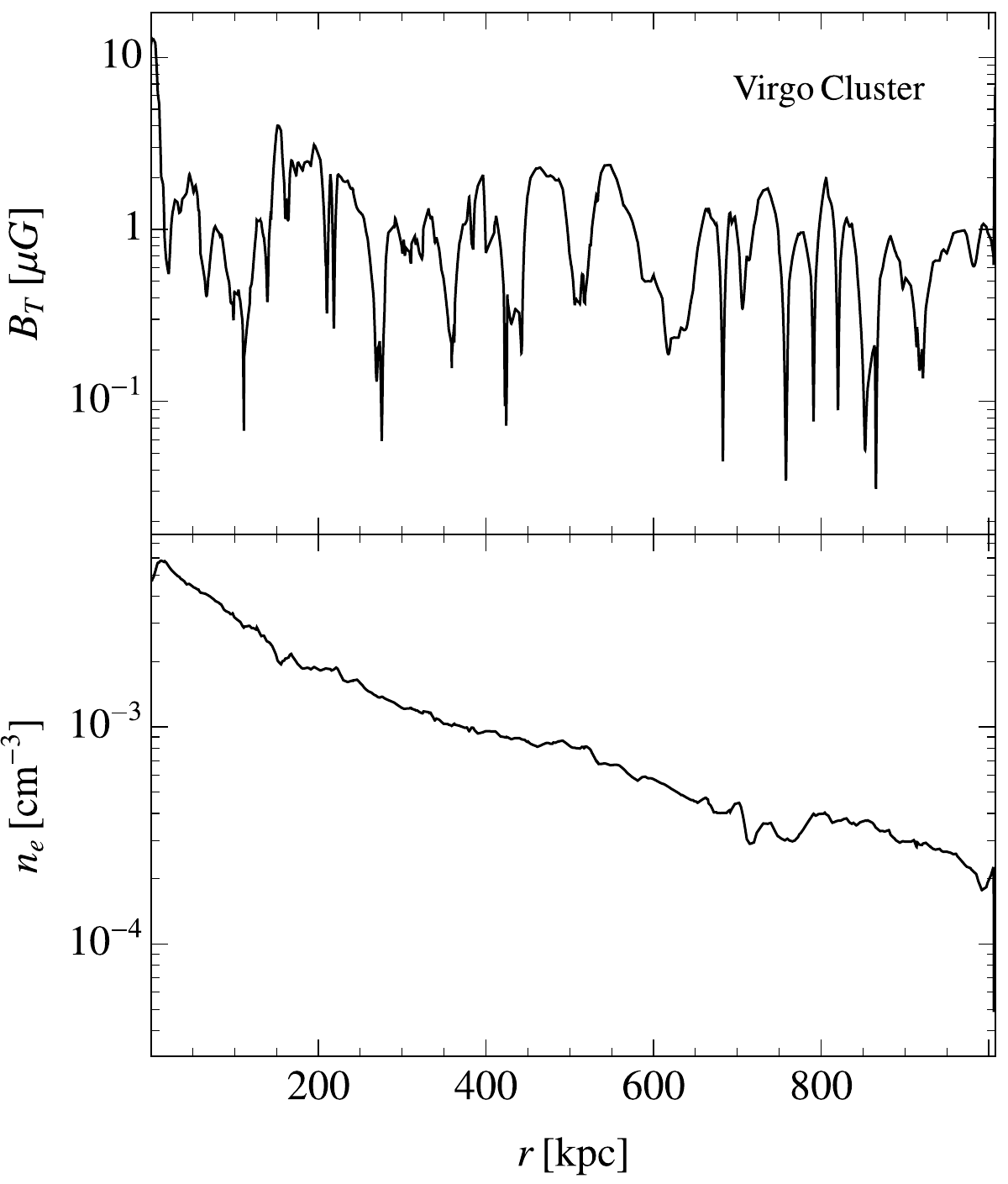}
\caption{Magnetic-field profiles (upper panels) and free-electron density profiles $n_e$ (lower panels) as a function of the distance from the galactic center for our fiducial line of sight. The left panel refers to M82 (solid line) and NGC\,6946 (dashed line), while results for the Virgo Cluster are reported in the right panel. 
}
\label{fig:plotA3}
\end{figure}

\begin{figure}[t!]
    \centering
    \includegraphics[width=0.55\columnwidth]{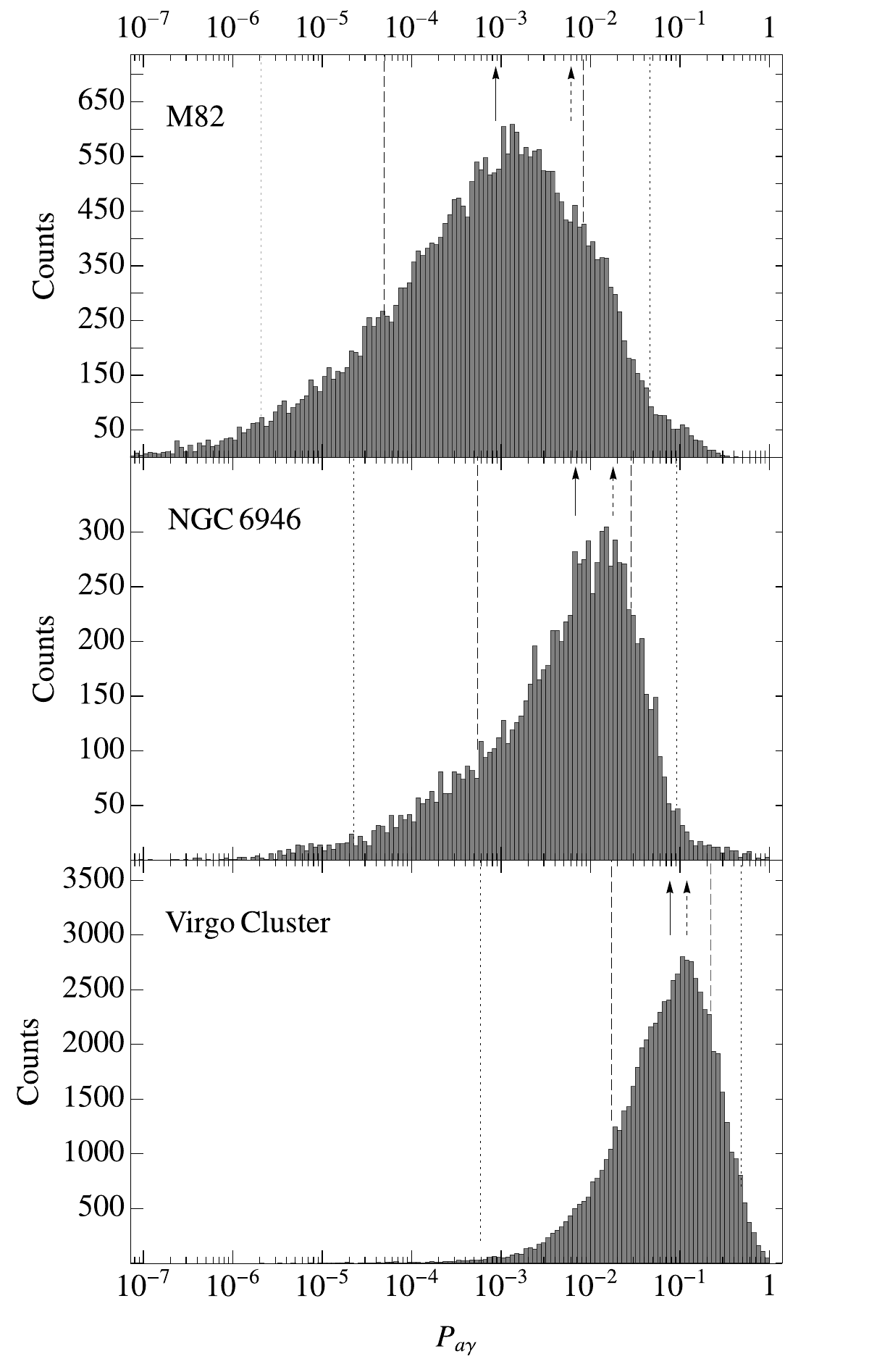}
    \caption{Distribution of conversion probabilities obtained by varying both the CCSN position and the observation direction within the various subhalo models for M82 (upper panel), NGC 6946 (middle panel), and the Virgo Cluster (lower panel). These results are obtained assuming $g_{a\gamma}=10^{-12}\,\text{GeV}^{-1}$ and $m_a \lesssim 10^{-10}\,\text{eV}$. The solid lines with the arrow indicates the median probability, while the dashed lines with arrow indicates the mean value. The dashed and dotted lines represent the $68\%$ and $95\%$ percentile intervals, respectively.} 
    \label{fig:plotA2}
\end{figure}

For each subhalo, we compute the conversion probabilities $P_{a\gamma}$ by solving the beam propagation equation, closely following the method of Ref.~\cite{Horns:2012kw}, assuming that ALPs are produced in a SN at the center of the host galaxy.
In  Fig.~\ref{fig:plotA2} we show the histograms of the conversion probability
distributions
obtained by varying both the CCSN position and the observation direction within each given subhalo employed for M82 (upper panel), NGC 6946 (middle panel) and Virgo Cluster (lower panel). For each distribution, we compute the meadian (solid line with an arrow in Fig.~\ref{fig:plotA2}) and the mean (dashed line with an arrow). Given the asymmetric nature of the conversion probability distributions, the median provides a robust estimate of the typical sensitivity limit, as it is less affected by extreme outliers in the magnetic-field realizations.

{We therefore adopt the median probability of each distribution, listed in Table~\ref{tab:Tab1}, to compute the sensitivities shown in Fig.~\ref{fig:fig1}. We define the line of sight associated to this median conversion probability as our \emph{fiducial} line of sight. {For illustrative purposes, in Fig.~\ref{fig:plotA3} we show the profile of the transverse magnetic field $B_T$ and free-electron density $n_e$ along our fiducial line of sight for M82 (left panel, solid line), for NGC 6946 (left panel, dashed line) and for Virgo Cluster (right panel, solid line)}. {We highlight that the structure of the magnetic field assumed for M82 is subject to sudden variations over correlation lengths shorter than NGC 6946. Thus, although the larger dimension of the assumed M82 subhalo, we expect the conversion probabilities to lose coherence at slightly higher ALP masses than NGC 6946 (see Fig.~\ref{fig:fig1}).}}

\begin{table}[t!]
    \centering
        \caption{Summary of the \textsc{IllustrisTNG} profiles used in this work. For each host galaxy, the Table lists the total stellar mass, the star-formation rate (SFR), the corresponding \textsc{IllustrisTNG} simulation volume, and the identifiers of the subhalos that best reproduce the observed properties of the respective environments.}
        \vspace{0.3cm}
   
    \begin{tabular}{c c c c c}
        \hline
        Host  & Total stellar mass& SFR & Simulation & Subhalos \\
        environment & ($M_{\odot}$) & ($M_{\odot}\,\mathrm{yr}^{-1}$) & volume &  \\
        \hline
        \hline
        M82 & $\sim 10^{10}$~\cite{Ning:2024eky} & $\sim 10$~\cite{Ning:2024eky} & 50 & 167408, 300907, 544408 \\
        NGC~6946 & $\sim10^{10.5}$~\cite{Lopez2023} & $\sim 5$~\cite{Tran2023} & 100 & 227576 \\
        Virgo Cluster & $\sim 10^{14}$~\cite{Ning:2024eky} & -- & 300 & 42631, 47315, 55060, 58081, 61682, 64929, 701461 \\
       \hline
    \end{tabular}
    
    \label{tab:A2}
\end{table}

\begin{figure}[t!]
    \centering
    \includegraphics[width=0.495\textwidth]{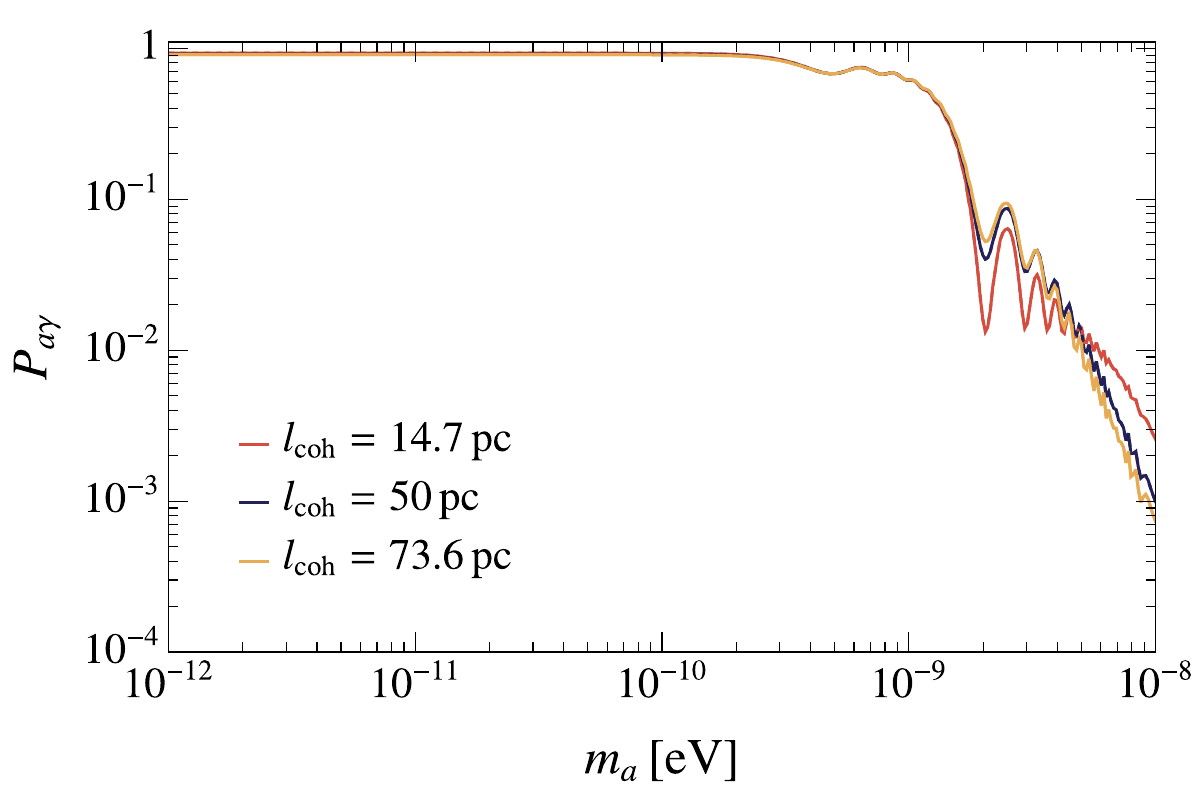}
    \includegraphics[width=0.495\textwidth]{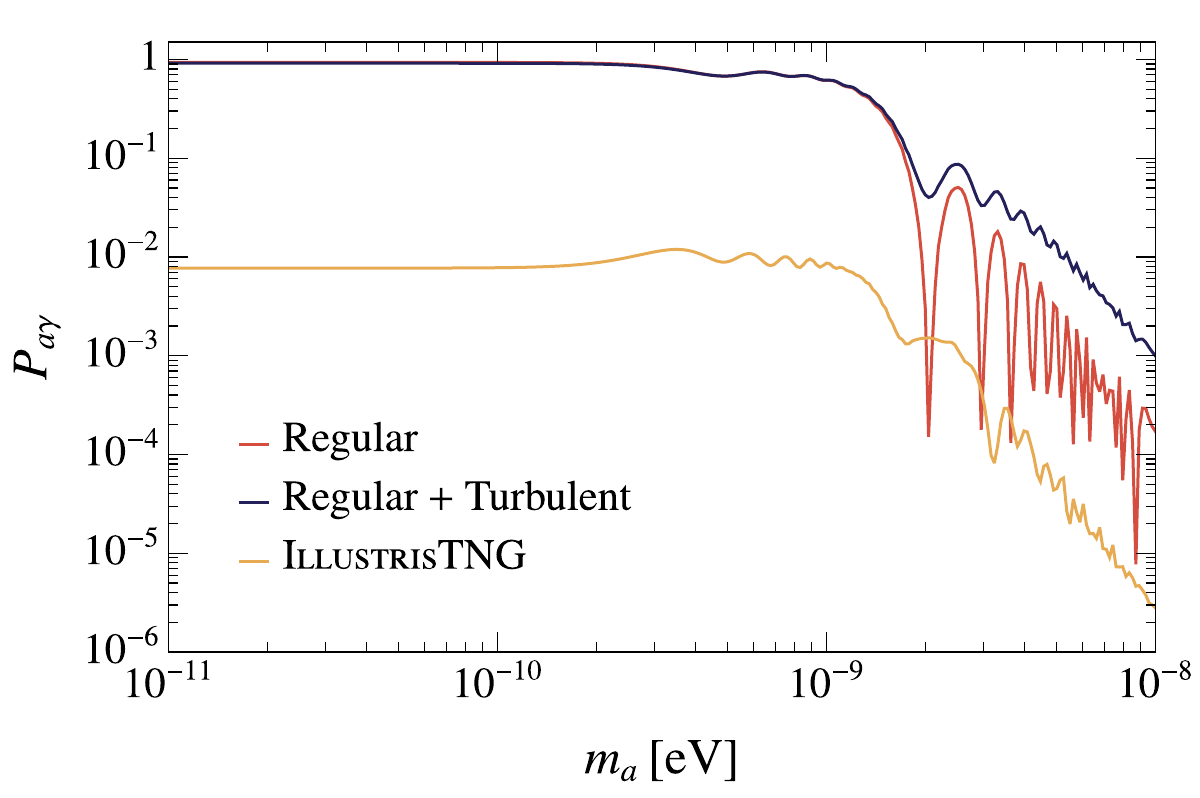}
    \caption{Axion-photon conversion probability $P_{a\gamma}$ as a function of the axion mass $m_a$ in M82. \textit{Left:} Realizations using the magnetic field parameters (regular and turbulent components) from Ref.~\cite{Lopez_Rodriguez_2021}. Each curve represents the conversion probability obtained by varying the coherence lengths. \textit{Right:} Comparison of $P_{a\gamma}$ for different modeling approaches: the regular component only (red curve) and the combined regular and turbulent components (black curve), both based on the magnetic field model of Ref.~\cite{Lopez_Rodriguez_2021}, together with our fiducial \texttt{IllustrisTNG} model (yellow curve).
    }
\label{fig:PagM82box}
\end{figure}

To assess the validity of the \textsc{IllustrisTNG} simulation results in the characterization of the magnetic fields of the considered astrophysical environments, we compare our findings with estimates derived from models available in the literature for the three different environments.  \\
For the starbust galaxy M82, we implemented a magnetic field model based on Ref.~\cite{Lopez_Rodriguez_2021}, which characterizes the field through a combination of mG-scale regular components and a stochastic turbulent component. We evaluated the axion-photon conversion probability by adopting the turbulence to-large-scale energy and coherence lengths reported in Ref.~\cite{Lopez_Rodriguez_2021}, testing different coherence lengths ranging from the typical Kolmogorov turbulence scale ($14.7$~pc), to larger values up to $\sim 50-75$~pc.
To avoid physically unreliable extrapolations where field data becomes sparse, we restricted the conversion region to $r\sim 10\,\rm kpc$. 
We find that the conversion probability remains consistently high  ($\sim 0.8$), as shown in the left panel of Fig.~\ref{fig:PagM82box}. Notably, this high efficiency is not strongly influenced by variation of coherence lengths or turbulence strength.

Furthermore, a comparison between this independent model and our fiducial \texttt{IllustrisTNG} approach is presented in the right panel of Fig.~\ref{fig:PagM82box}. The results indicate that the $\mathrm{mG}$-scale fields present in the starburst core could improve axion sensitivity limits by approximately an order of magnitude. This external validation confirms that the modeling used in this work, as also noted in  Ref.~\cite{Ning:2024eky}, represents a conservative estimate for the case of M82.

For NGC~6946 we used the magnetic field model presented in Ref.~\cite{Khademi_2023}, which consists of a spiral component confined to the galactic plane and an X-shaped poloidal component.
However, since no explicit model or numerical values are provided for the X-shaped component, only the spiral magnetic field was included in our analysis.
This model features a minor uncertainty associated with the maximum field strength along the spiral arms, which can vary between $8~\mu\mathrm{G}$ and $10~\mu\mathrm{G}$. 
To model the electron number density, the average values of $n_e$ reported in Ref.~\cite{Ehle1993} are adopted (Tab.~8, model~3).
Using these inputs, the conversion probability in the massless regime is found to be $P_{a\gamma} = 6.8\times 10^{-3}$, which is consistent with the value reported in Table~\ref{tab:Tab1}, obtained using \textsc{IllustrisTNG}.\\

To validate our modeling of the Virgo Cluster, we implemented the magnetic field and electron density profile described in Ref.~\cite{Marsh:2017yvc}, hereafter referred as Random Field (RF) model. In this framework, the magnetic field is formulated as a 1D radial profile, with its orientation and coherence length drawn from a probability distribution. To ensure a one-to-one comparison with our analysis, we extend this 1D prescription to construct a full 3D configuration. Specifically, we build a 3D realization by generating independent 1D radial profiles along various angular directions outward from the cluster center, effectively assembling these independent radial LoSs into a complete spherical geometry. Within this 3D model, we evaluated the conversion probability for $10^2$ random SN position, testing $10^{2}$ random orientations for each position.

\begin{figure}[t!]
    \centering
    \includegraphics[width=0.495\textwidth]{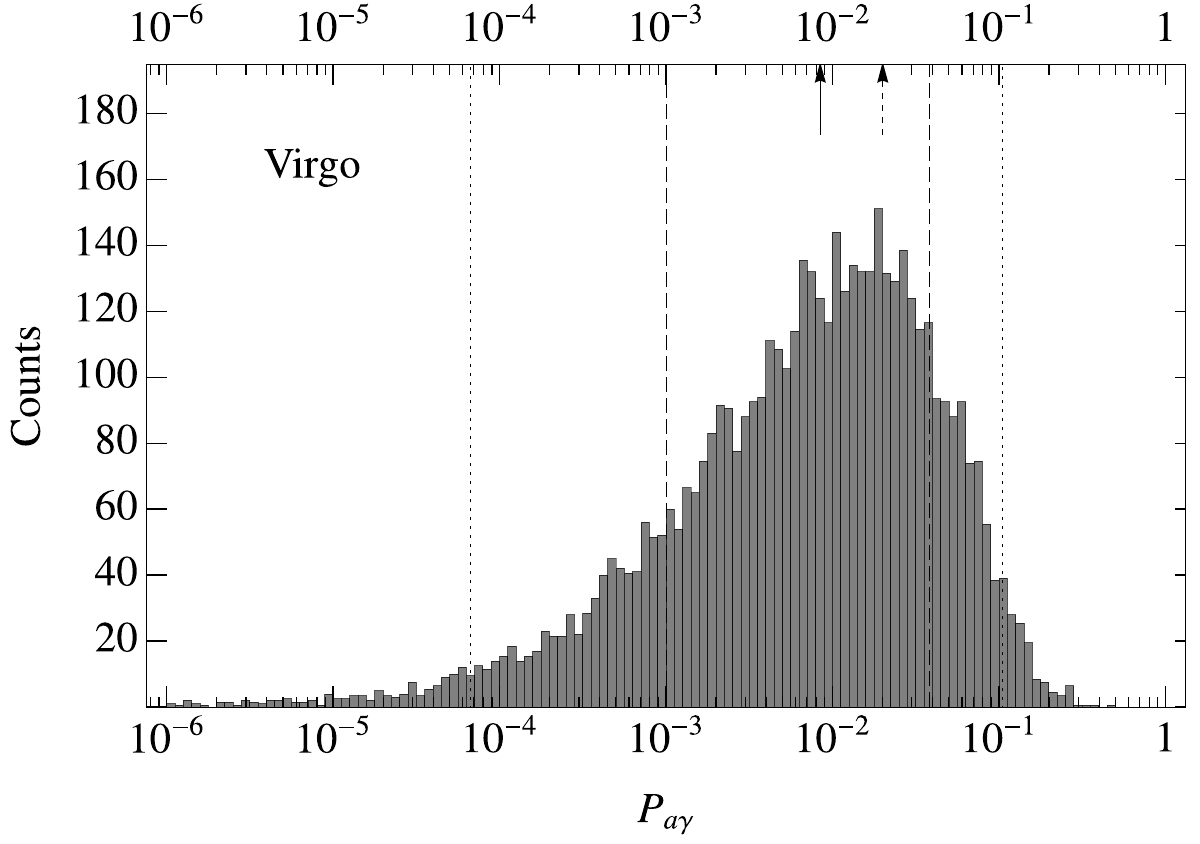}
    \includegraphics[width=0.495\textwidth]{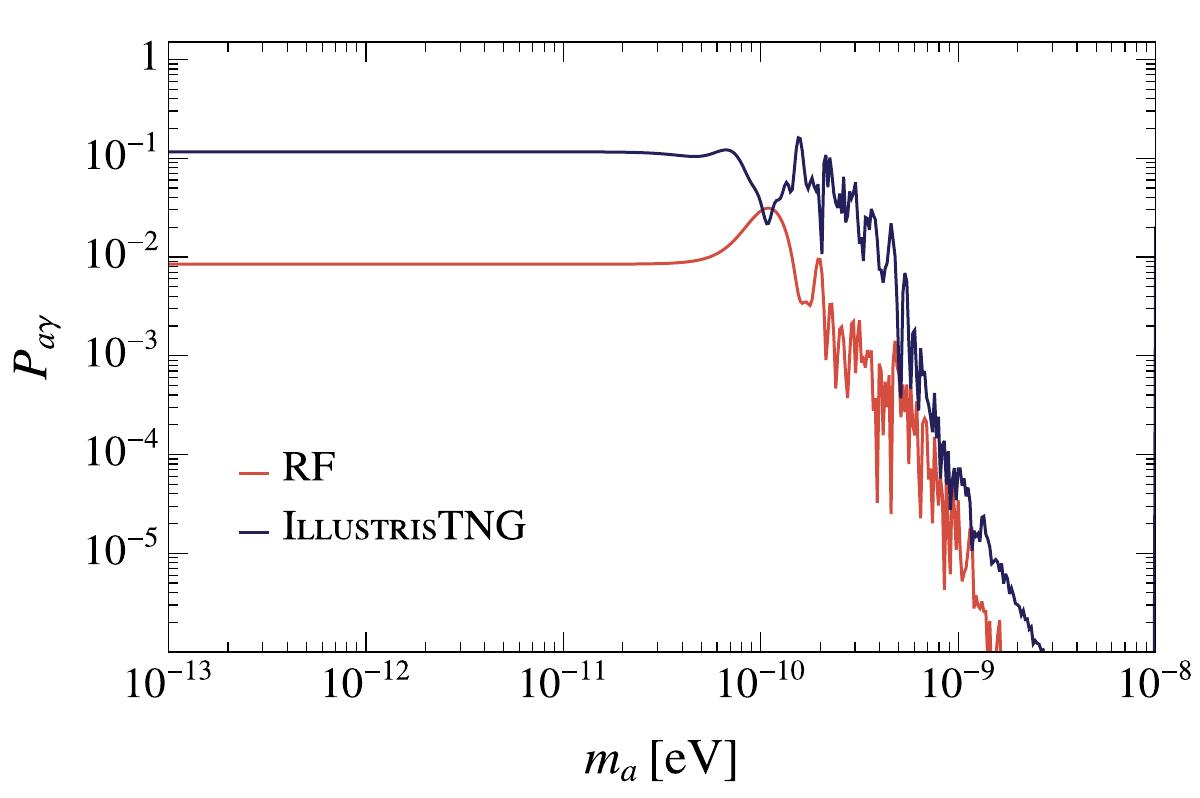}
     \caption{Axion-photon conversion in the Virgo Cluster. \textit{Left:}Distribution of conversion probabilities for $E = 50~\mathrm{MeV}$, obtained using the Random Field (RF) model of the Virgo Cluster as described in  Ref.~\cite{Marsh:2017yvc}. 
     The solid lines with the arrow indicates the median probability, while the dashed lines with arrow indicates the mean value. Right: Comparison of $P_{a\gamma}$ for different modeling approaches: the median (red curve) conversion probability obtained using the RF magnetic field model, shown alongside our fiducial \texttt{IllustrisTNG} model (black curve).}
     \label{fig:NewM87prob}
\end{figure}

The resulting distribution is shown in the left panel of Fig.~\ref{fig:NewM87prob}. From this ensemble, we extracted the mean and median probability conversions.

The final conversion probabilities as a function of axion masses are presented in the right panel of Fig.~\ref{fig:NewM87prob} and compared with our fiducial \texttt{IllustrisTNG} model. We find that the conversion probability yielded by the RF model in the massless limit are approximately one order of magnitude weaker than those obtained using the \texttt{IllustrisTNG} configuration. As axion-photon conversion probability scales with the field correlation length as $P_{a\gamma}\sim l_{\rm corr}$ , magnetic fields which are correlated on shorter scales led to less efficient conversions. 
For the RF mode, taking the average value of the coherence length distribution shown in Ref.~\cite{Marsh:2017yvc} yields a typical correlation length of $l_{\rm corr}\sim3~\,\rm{kpc}$. Similarly, based on a statistical analysis of different LoS across the various subhalos, we find a typical correlation length of $l_{\rm corr}\sim10~\,\rm{kpc}$ for the \texttt{IllustrisTNG} model. This difference in correlation lengths justifies the difference in conversion probabilities.

Uncertainties in the adopted magnetic-field model then translate into a factor of $\sim3$ uncertainty in the constraints derived in the main text.


To gain some physical intuition,  it is also useful to study approximate analytical solutions.
As shown in Ref.~\cite{Calore:2023srn}, {in the massless case} 
 the conversion probability becomes energy independent, and for $\Delta_{a\gamma} L <1$ reduces to
\begin{equation}
P_{a\gamma} \simeq (\Delta_{a \gamma}L)^2 \,\ ,
\label{eq:enindep}
\end{equation}
where $L$ is the size of the magnetic domain, and
\begin{eqnarray}  
\Delta_{a\gamma}\!\!\! &\simeq &\!\!\! 1.5\times10^{-3} \left(\frac{g_{a\gamma}}{10^{-12}\,\GeV^{-1}} \right)
\left(\frac{\langle B_{T}\rangle}{10^{-6}\,\rm G}\right) \kpc^{-1} \nonumber\,,  \\
\end{eqnarray}
with $\langle B_T \rangle$ the average of the transverse component of the magnetic field along the line of sight. For an ALP moving in the $z$ direction, this is given by 
\begin{equation}
    \left< B_{T} \right> = \frac{1}{L} 
  \sqrt{
     \left| \int_{0}^{L} \! dz \, B_{x}(z) \right|^{2} 
        + 
        \left| \int_{0}^{L} \! dz \, B_{y}(z) \right|^{2}
    } .
\end{equation}
The values of the parameters entering Eq.~(\ref{eq:enindep}) for the three considered galaxies are reported in Table~\ref{tab:Tab1}. {In the massless limit ($m_a \ll 10^{-9}$~eV), the probability from Eq.~\eqref{eq:enindep} reported in Tab.~\ref{tab:Tab1} agrees with the result obtained using the method of Ref.~\cite{Horns:2012kw}.}

\begin{table*}[t!]
\caption{Observed supernovae for the period 2008--2025 in M82, NGC 6946 and in Virgo Cluster. The table reports also the average transverse magnetic field $\langle B_{T}\rangle$ along the {fiducial} line of sight, the extension of the conversion path} $L$,  {its distance from Earth $d$, the median ALP-photon conversion probability $P_{a\gamma}$ for massless ALPs with 
$g_{a\gamma}=10^{-12}\,\GeV^{-1}$, and the sensitivity to $g_{ap} \times g_{a\gamma}$ in the massless case.}
    \centering
    \begin{tabular}{c c c c c c c}
    \hline
 Host     &Observed SNe & $\langle B_{T} \rangle$& $L$&$d$& $P_{a\gamma}$ & $g_{ap} \times g_{a\gamma}$\\
  Environment    & (2008-2025) & ($\mu$G) & (kpc) & (Mpc) & & {($\times 10^{-24}\,\ \textrm{GeV}^{-1}$)} \\
      \hline
      \hline
      M82& SN 2008iz&$1.4$  & $13.9$& $3.5$ & $8.8\times 10^{-4}$&${2.2}$  \\
      NGC 6946&SN 2008S,SN 2017eaw & $ 5.8$  & $9.3$&$5.9$& $6.8 \times 10^{-3}$& ${1.3}$  \\
      Virgo Cluster $ $  & SN 2012cc, SN 2014L, SN 2020oi, SN 2019ehk & $1.1$  & $10^{3}$&$16.4$& $7.6\times 10^{-2}$& ${1.1}$  \\
      \hline
    \end{tabular}
    \label{tab:Tab1}
\end{table*}

\section{Sensitivities in ALP-induced gamma-ray signals}
\label{App:sensitivities}

\begin{figure}[t]
    \centering
    \includegraphics[width=0.5\linewidth]{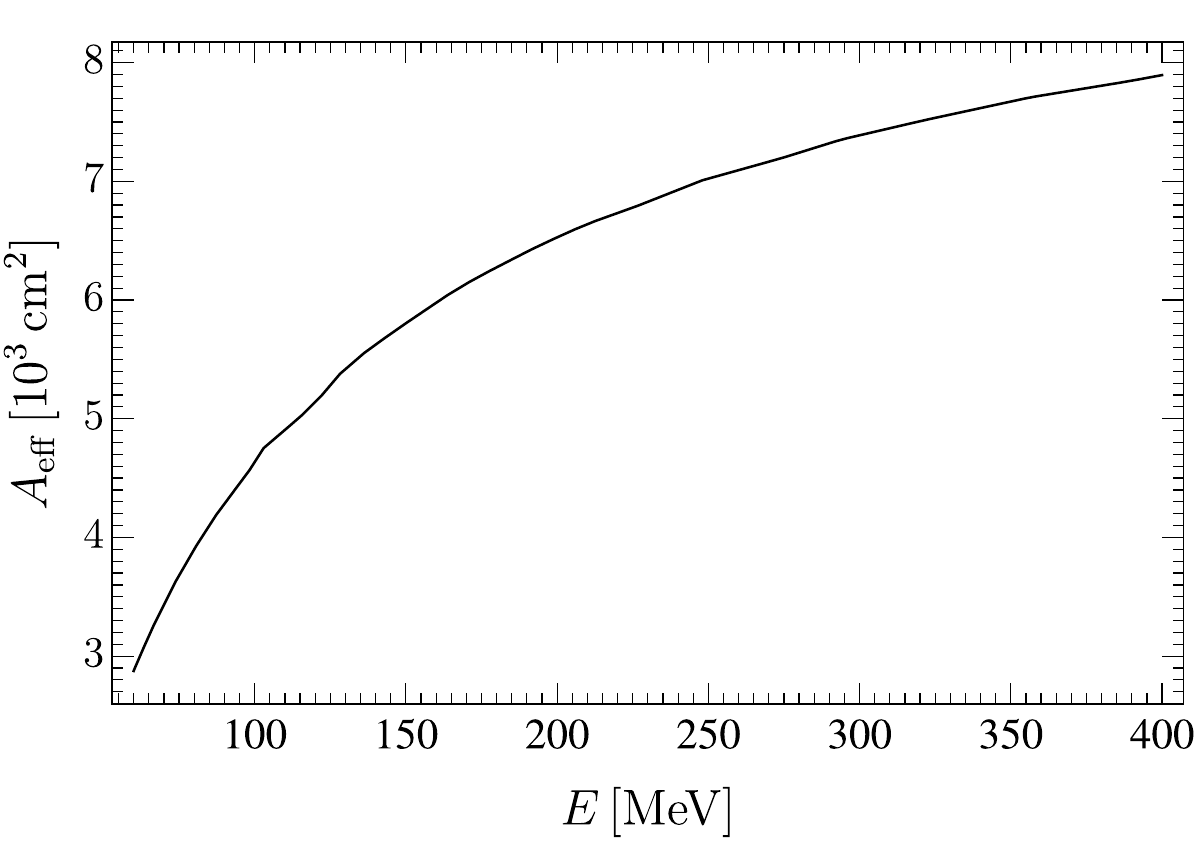}
     \caption{{On-axis effective area of \textit{Fermi}-LAT for transient events (dotted curve in Fig. 14 of Ref.~\cite{Fermi-LAT:2009ihh}).}}
    \label{fig:plotA4}
\end{figure}

We characterize the detectability of a gamma-ray burst induced by SN ALP conversions,  assuming a telescope with the same performances of  {\it Fermi}-LAT~\cite{Bruel:2018lac}. We compute the number of expected gamma-ray events as
\begin{equation}
\label{eq:nev}
    N_{\mathrm{ev}} = 
    \int_{E_{\mathrm{min}}}^{E_{\mathrm{max}}} dE \,
    \frac{d\phi_\gamma}{dE} \, 
    A_{\mathrm{eff}}(E) \, ,
\end{equation}
where ${d\phi_\gamma}/{dE}$ is the ALP-induced photon flux reaching the Earth, 
$A_{\mathrm{eff}}$ is the \textit{Fermi}-LAT effective area, and 
$[E_{\mathrm{min}}, E_{\mathrm{max}}] = [60, 250]~\mathrm{MeV}$ is our considered energy window, above the energy threshold of {\it Fermi}-LAT. 
The photon flux produced by ALP-photon conversion in the host galaxy magnetic field and reaching the Earth is given by
\begin{equation}
    \frac{d\phi_{\gamma}}{dE}(E,L) = 
    \frac{1}{4\pi d^{2}} \,
    \frac{dN}{dE}(E) \,
    P_{a\gamma}(E,L) \,\ ,
    \label{eq:photonflux}
\end{equation}
where $dN/dE$ is the ALP production spectrum (see SM~B), 
$P_{a\gamma}(E,L)$ is the ALP--photon conversion probability, 
and $d$ is the distance to the host galaxy (fifth column of Table~\ref{tab:Tab1}).  
The resulting photon flux from M82 for $g_{a\gamma}=10^{-12}$~GeV$^{-1}$ is shown in Fig.~\ref{fig:plotA1}. The effective area $A_{\mathrm{eff}}(E)$ used in this analysis is the one of Fermi-LAT for normal incident photons~\cite{Bruel:2018lac} (see Fig.~\ref{fig:plotA4}). It is reasonable to use the on-axis effective area, as GALAXIS's all sky coverage implies that most photons arrive with small incident angles. Moreover, Ref.~\cite{Fermi-LAT:2013jgq} shows that the incident angle has only a minor impact on the effective area~\cite{Candon:2025sdm}.
In order to describe the emission from the background sources and to extract the expected sensitivity from a given SN, 
 one should derive a model for a region of interest  (see Ref.~\cite{Meyer:2020vzy}). In our simplified analysis we assume that we are able to extract the gamma-ray emission signal from this region and therefore we compare the ALP-induced signal with the only diffuse gamma-ray background. Then,
the background event rate is computed as

\begin{equation}
  R_{\rm{bkg}} = \Omega \times \int_{E_{\rm min}}^{E_{\rm max}} dE\, \frac{d\phi_{\gamma,\rm{bkg}}}{dE}\, A_{\rm eff}(E)\ ,\, 
  \label{eq:background}
\end{equation}
where 
\begin{equation}
\Omega =2\pi \left(1 - \cos{\delta \theta}\right),
\end{equation}
is the solid angle corresponding to the angular resolution ${\delta \theta}$. For our analysis we adopt an angular resolution of $3.5^\circ$~\cite{Fermi:LATSpecifications}. The quantity $d\phi_{\gamma,\mathrm{bkg}}/dE$ is the diffuse gamma-ray background reported in {\tt iso\_P8R3\_SOURCE\_V3\_v1.txt} from Ref.~\cite{Fermi:BackgroundModels}. {The resulting background event rate is $R_{\rm bkg} =7.7\times 10^{-4}\,\rm{s}^{-1}.$} The number of background events is then $N_{\rm bkg} =R_{\rm bkg}\times \Delta t$, with $\Delta t$ the observation time window. We consider three cases, namely $\Delta t=1$~d and $\Delta t=1$~h, determined by the accuracy in the determination of the SN core-collapse time provided by optical survey, as well as $\Delta t = 10$~s, corresponding to the ALP burst duration. The latter can be adopted only if a GW trigger determines the core-collapse time.

We find  $N_{\mathrm{bkg}} \simeq 67$ 
for $\Delta t=1$ d, indicating a relatively large background compared with the ALP-induced gamma-ray expected for the explored couplings. Following the procedure described in 
Ref.~\cite{ParticleDataGroup:2024cfk}, under these conditions 
 we evaluate the $95\%$ CL sensitivity as
\begin{equation}
   N_{\rm ev}> 1.64 \sqrt{N_{\rm bkg}} \,\ .
\end{equation}
For  $\Delta t= 1~$h  we find $N_{\rm bkg} \simeq 3$, while for $\Delta t = 10$~s we obtain $N_{\rm bkg} \simeq 0$, corresponding to a low-background and free background regime, respectively. In these situations, we adopt the Feldman–Cousins prescription~\cite{Feldman:1997qc}, which is appropriate for the low-count limit. For these values of the background events, the $95\%$~CL sensitivity corresponds to requiring $N_{\rm ev} \gtrsim 5$ for $\Delta t=1\,$h, and $N_{\rm ev} \gtrsim 3$ for $\Delta t=10$~s~\cite{Feldman:1997qc}.

The background estimation has been validated through two independent assessments to ensure the modeling remains both consistent and conservative. By implementing a more rigorous $4\sigma$ significance requirement ($N=4\sqrt{N_{\rm{bkg}}}$), we confirm that for a single SN event, the $\Delta t = 10$~s and  $\Delta t=1$~h windows  remain effectively background free. This justifies the use of the Feldman–Cousins statistical framework for calculating discovery thresholds in these regimes. On the other hand, the sensitivity for the $\Delta t=1$ d window shifts by a factor of approximately $1.2$ under this criterion. Furthermore, an investigation into the specific environments of the potential hosts using the \emph{Fermi}-LAT 4FGL-DRA catalog~\cite{Ballet:2026kpq} reveals that the emission from resolved sources is consistently lower than the isotropic background rate. This confirms that treating the isotropic background as the dominant noise source is a robust and conservative approach.

If multiple SN explosions are observed, the combined (stacked) analysis improves the sensitivity to the product $g_{ap}\times g_{a\gamma}$. Since both the signal and background scale linearly with the number of stacked SNe $N_{\rm SN}$, in the background-dominated regime we require the signal to background ratio to satisfy
\begin{equation}
    N_{\rm SN} N_{\rm ev}> 1.64 {\sqrt{N_{\rm SN} N_{\rm bkg}}} \,.
\end{equation}
Since $N_{\rm ev}\propto g_{ap}^2\times g_{a\gamma}^2$, the sensitivity scales as
\begin{equation}
    g_{ap}\times g_{a\gamma} \propto N_{\rm SN}^{-1/4}\,.
\end{equation}

{For observations with $\Delta t=10$~s the background remains smaller than one event for $N_{\rm{SN}} \lesssim 200$.} In this case only the number of events scales with the number of SNe, and the corresponding sensitivity behaves as
\begin{equation}
    g_{ap} \times g_{a\gamma} \propto N_{\rm SN}^{-1/2}\,.
\end{equation}

{Being in a low-background regime, for $\Delta t=1$~h  the $95\%$ CL sensitivity is determined using the thresholds tabulated in Table XII of Ref.~\cite{Feldman:1997qc}. This results in a sensitivity scaling that lies between the background-dominated and background-free regimes.}

\section{Uncertainties in ALP-induced gamma-ray signals}
\label{app:uncertainties}

The sensitivities to $g_{ap} \times g_{a\gamma}$ \emph{vs} $m_a$ shown in Fig.~\ref{fig:fig1} are obtained with our fiducial line of sight, i.e. using the subhalo and line of sight which yield the median value of the distributions presented before, and taking $\Delta t=10$~s. 

\begin{figure}[t!]
    \centering
    \includegraphics[width=0.7\linewidth]{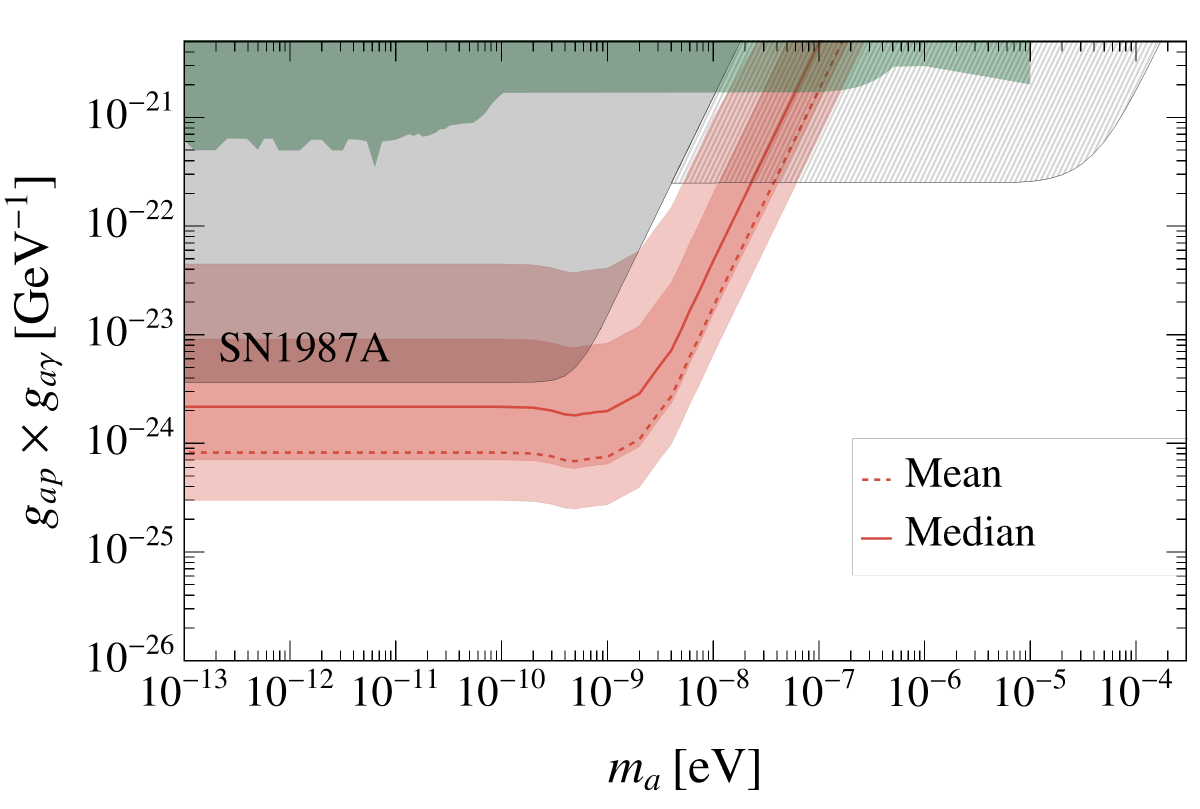}
    \caption{Uncertainties in the sensitivity curves obtained by varying both the CCSN position and the observation direction in M82. The boundaries of the dark (light) red shaded band correspond to the lines of sight providing conversion probabilities enclosing the the $68\%$ ($95\%$) percentile intervals around the central value of the distribution. The solid (dashed) red line indicates to the sensitivity obtained with the median (mean) probability conversion shown in Fig.~\ref{fig:plotA2}.}
    \label{fig:plotsensnewM82100idr100pos}
\end{figure}

As previously discussed, confidence intervals are also extracted. Figure~\ref{fig:plotsensnewM82100idr100pos} shows the resulting uncertainty band for M82, obtained by evaluating the sensitivities using the subhalos and lines of sight corresponding to the extrema of the $68\%$ and $95\%$ confidence-level intervals. Due to the large variability of magnetic field strengths along the different
subhalo models and various line of sights considered in the analysis, the uncertainty bands over $g_{ap}\times g_{a\gamma}$ may extend over more than $1$ order of magnitude.

We further compare in Fig.~\ref{fig:plotA7} the sensitivities obtained using the median (solid line) and the mean (dashed) conversion probabilities. This comparison shows that the median yields sensitivities weaker by a factor of a few. While the median provides a more robust statistical estimator in the presence of highly asymmetric distributions, the mean values align more closely with probabilities obtained by using existing modeling in literature as shown in Sec.~C. This highlights that our main results in Fig.~\ref{fig:fig1}, based on the median, represent conservative estimates of the sensitivity.

\begin{figure}[t]
    \centering
    \includegraphics[width=0.7\linewidth]{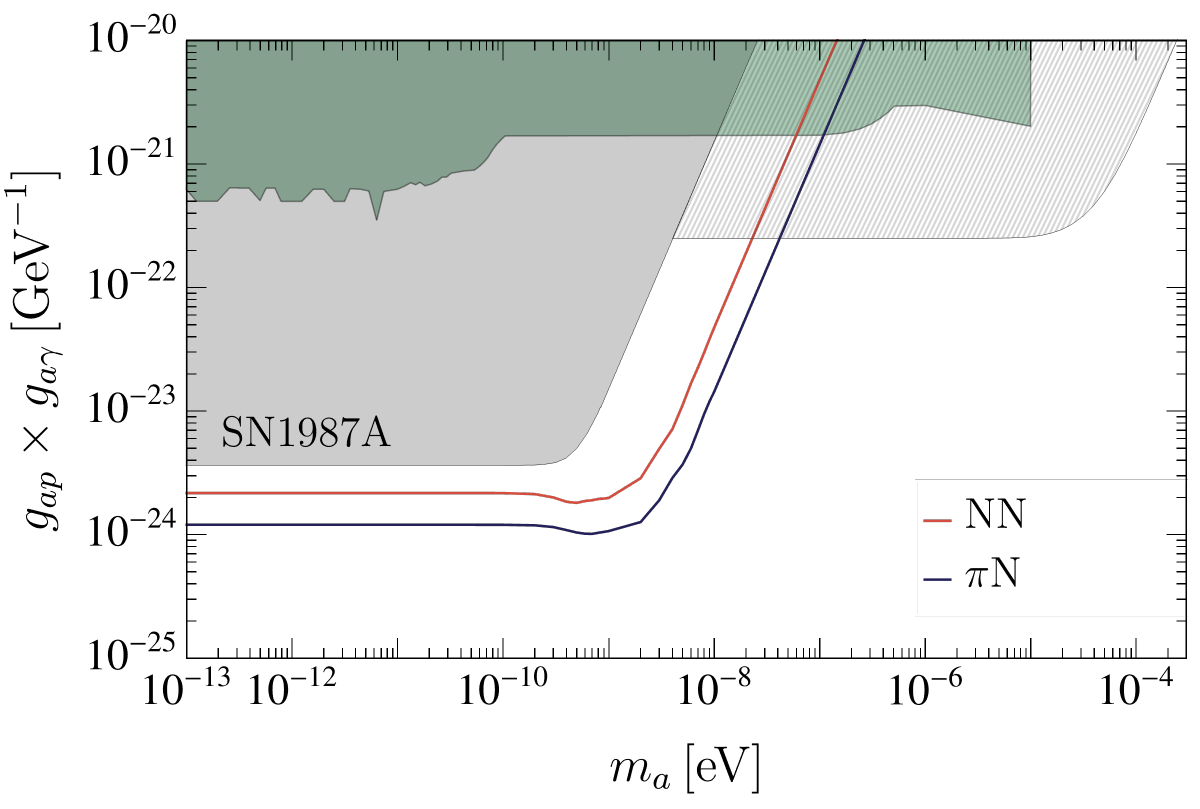}
    \caption{{Sensitivities at $95\%$ CL for the fiducial M82 line of sight by considering ALP production via NN bremsstrahlung only (red) and by including also the contribution from pionic processes (blue). See text for details. The color scheme for other constraints is the same as the one used in Fig.~\ref{fig:fig1}.}
    }
    \label{fig:plotA6}
\end{figure}

A further source of uncertainty arises from the inclusion of pions in the production mechanism.
Indeed, the role of pions in supernova cores remains uncertain, as they are often omitted in SN models. Earlier virial-expansion studies suggested that 
 could reach abundances comparable to muons, growing with density and potentially affecting axion production~\cite{Fore:2019wib,Lella:2022uwi}. However, more recent chiral-effective-theory calculations find strongly increased in-medium pion energies, which would suppress pion abundance~\cite{Fore:2023gwv}
(see the discussion of  Ref.~\cite{Fiorillo:2025gnd}).
{In order to quantify the impact of pionic processes,  in 
Fig.~\ref{fig:plotA6} we show how the sensitivity plot for  M82 changes if one includes these latter. 
We find an increase in sensitivity due to presence of pion processes by a factor $\sim 2$.}

\section{ALPs coupling to photons only}
\label{app:photon coupling}

\begin{figure}
    \centering
    \includegraphics[width=0.7\linewidth]{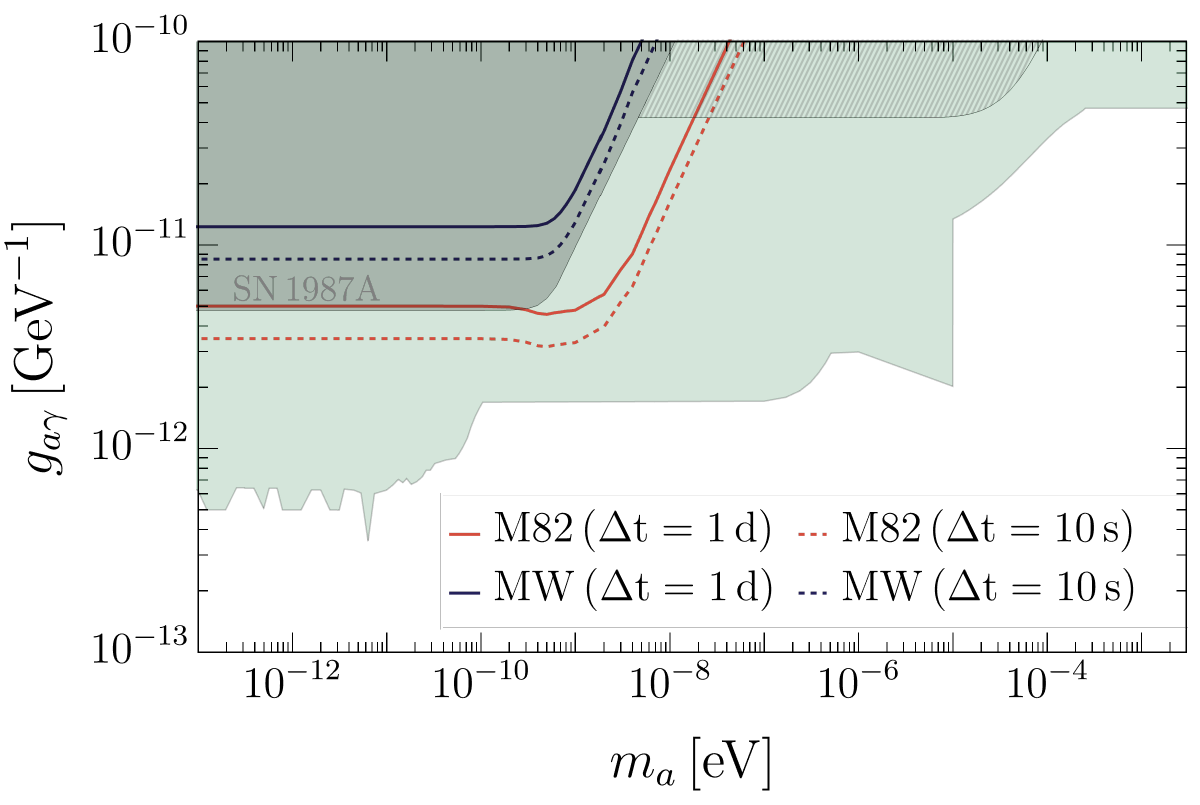}
    \caption{{Sensitivities at 95 \% CL in the plane $g_{a\gamma}$ \emph{vs} $m_a$  for an extragalactic SN located in M82 assuming ALP-photon conversions in the Milky-Way (blue curves) and along a fiducial line of sight through the host-galaxy magnetic field (red curves). In this plot, we set ALP-nucleon $g_{aN}=0$, so that ALP production in the SN core is triggered solely by the Primakoff process. Solid lines refer to $\Delta t=1$~d and dashed lines to $\Delta t=10$~s.}
    }
    \label{fig:plotA7}
\end{figure}

{ Finally, in order to compare our results with previous literature~\cite{Meyer:2020vzy}, we consider the case of ALPs coupled with only photons and converting into gamma rays only in the Milky-Way magnetic field. We take as reference an extragalactic SN in M82. In case of only photon-ALP coupling, ALP would be produced in a SN core via Primakoff process~\cite{Dicus:1978fp,Raffelt:1985nk}.
In this situation,  we use as SN ALP flux the one presented in Ref.~\cite{Fiorillo:2025gnd}  for the one-zone model for a cold SN. We calculate ALP-photon conversions in the Milky Way following 
 Ref.~\cite{Calore:2023srn} {and find  
 $P_{a\gamma}\sim {\mathcal O (10^{-5})}$ for low-mass ALPs and $g_{a\gamma}=10^{-12}$~GeV$^{-1}$.} In Fig.~\ref{fig:plotA7}, we show the $95\%$ CL sensitivities in the $g_{a\gamma}$ {\emph vs} $m_a$ plane for the situation described above, using an observation-time window 
$\Delta t=1~$d. We find a sensivity to 
$g_{a\gamma} \gtrsim {1.2} \times 10^{-11}$~ GeV$^{-1}$ for $m_a \ll 10^{-9}$~eV, which is a factor {$\sim 2.5$} worse in comparison with the corresponding SN 1987A bound. 
These results are consistent with those reported in Ref.~\cite{Meyer:2020vzy}. {For example, for SN~2017ein at a distance of $d \approx 11.6$~Mpc (see Fig.~1 of \cite{Meyer:2020vzy}), Ref.~\cite{Meyer:2020vzy} obtained a bound of $2\times 10^{-11}$~ GeV$^{-1}$. After rescaling for the different distance assumed in our analysis ($d = 3.5$~Mpc for M82), this agrees well with our result.}
We realize that the improvement  in the presence of $\Delta t=10$~s is a factor {1.4}. 
For comparison if we include also conversions in the magnetic field of the host environment we find
a sensitivity to $g_{a\gamma} \gtrsim {5} \times 10^{-12}$~ GeV$^{-1}$ for $\Delta t=1~$d and 
$g_{a\gamma} \gtrsim {3.5} \times 10^{-12}$~ GeV$^{-1}$
for $\Delta t=10$~s, with an improvement of a factor $3.4$ with respect to conversions in the only Milky-Way, {but still excluded by other astrophysical constraints.}} 

\end{document}